\documentclass[aps,twocolumn]{revtex4}

\usepackage{graphicx}

\usepackage{bm}
\usepackage{graphicx}
\usepackage{amsmath}
\usepackage{amssymb}
\usepackage[dvips]{color}

\begin{document}

\title{Associative memory of phase-coded spatiotemporal patterns  in
leaky Integrate and Fire networks
}

\author{Silvia Scarpetta$^{1,2}$  ,              Ferdinando Giacco$^{3}$   
}

\affiliation{
$^{1}$ Dept. of  Physics  ``E. R.\ Caianiello'',  University of Salerno,
84084 Fisciano (SA), IT\\   
$^{2}$INFN  Unita' di Napoli Gruppo coll. di Salerno,
IT\\
$^{3}$              Dep. of Environmental Sciences, Second University of Naples,
81100
Caserta, Italy
}

\date{Received: date / Accepted: date}


\begin{abstract}
We study the collective dynamics of a  Leaky Integrate and Fire  network 
in which precise relative phase relationship  of spikes among neurons are
stored, as attractors of
the dynamics, and selectively replayed at different time scales.
Using an STDP-based learning process, we store in the connectivity  several
phase-coded spike patterns, 
and we find that, depending on the excitability of the network, different
working regimes are possible,
with transient or persistent replay activity induced by a brief signal. 
We introduce an order parameter to evaluate the similarity  between
stored and recalled phase-coded pattern, and
measure the storage capacity.
Modulation of spiking thresholds during replay changes
 the frequency of the collective oscillation or the number of spikes per
cycle,
keeping preserved the phases relationship.  This allows a coding scheme in
which phase, rate and frequency are dissociable. 
Robustness with respect to noise and heterogeneity of neurons parameters is
studied, showing  that, since dynamics 
is a retrieval process, neurons preserve stable precise phase relationship
among units, keeping a unique frequency of oscillation, 
even in noisy conditions and with heterogeneity of internal parameters of
the units. 
\end{abstract}

\maketitle
\section{Intro}
\label{sec_intro} 

It has been hypothesized that,   in many areas of the brain, having different brain functionality,
repeatable precise spatiotemporal patterns of spikes play a crucial role in coding and storage of information.
Temporally structured replay of  spatiotemporal patterns have been observed to occur during sleep, both in the cortex and hippocampus 
\cite{buz1,JiWilson,Bruce,Bruce2009}, and it has been hypothesized that this replay may subserve memory consolidation. 
The sequential reactivation of hippocampal place cells, corresponding to previously experienced
behavioral trajectories, has been observed also in the awake state (awake replay) \cite{NatureN2007,extended,ripples,2011}, namely  during periods of relative immobility. 
Awake replay may reflect trajectories through either the current environment or previously, spatially remote, visited environments. A possible interpretation is   
that spatiotemporal patterns,  stored in the  plastic synaptic connections of
hippocampus, are  retrieved when a cue activates the
emergence of a stored pattern, allowing these patterns to be replayed and then consolidated in distributed
circuits beyond the hippocampus \cite{2011}.  
Cross-correlogram analysis  revealed that  in prefrontal  cortex the time scale of reactivation of firing patterns during post-behavioral sleep was compressed five- to eightfold  relative to waking state \cite{Bruce,Bruce2011}, a similar compression effect
may also be seen in primary visual cortex\cite{JiWilson}.
Internally generated spatiotemporal patterns have also been observed in the rat
 hippocampus during the delay period of a memory task, showing that the emergence
of consistent pattern of activity  may be a way to maintain important
information during a delay in a task \cite{buz}.

Among  repeating patterns of spikes a central role is played by  phase-coded patterns 
 \cite{MillerPNAS,panzeri-kayser,panzeri-phase1,montemurro}, i.e. patterns with precise relative phases of the
spikes of neurons participating to a collective oscillation, or precise phases of spikes relatively
to the ongoing oscillation. 

First experimental evidence of the importance of spike phases in neural coding was observed in
experiments on theta phase precession in rat's place cells \cite{Okeefe,Burgess},
showing that spike phase is correlated with rat's position.
Recently,  the functional role of oscillations in the hippocampal-entorinal cortex 
circuit for path-integration 
has been deeply investigated \cite{Okeefe,Lengyel,Burgess,Bruce,GeislerPNAS,Moser}, showing that place cells and grid cells 
form a map in which precise phase
 relationship among units plays a central role.
In particular  it has been shown\cite{Burgess-2,Blair2008,Welday2011}  that
both spatial tuning and phase-precession properties of place cells
can arise when one has interference among oscillatory
 cells with precise phase relationship and  velocity-modulated frequency.

Further evidence of phase coding comes 
from the experiments on spike-phase coding of natural stimuli in auditory and visual primary cortex  
\cite{panzeri-phase1,panzeri-kayser}, and
from experiments 
on short-term memory of multiple objects in prefrontal cortices of monkeys \cite{MillerPNAS}.

These experimental works   support the hypothesis  that collective oscillations may underlie a 
phase dependent neural coding and an associative memory behavior which is able to recognize the phase coded patterns.

The importance of precise timing relationships among neurons, which may carry information to be stored,
is supported also by the evidence that precise timing of few milliseconds is able to change the sign of synaptic plasticity.
The dependence of synaptic modification on the precise timing and order of pre- and post-synaptic spiking has been demonstrated in a variety of neural circuits of different species.
Many experiments show that a synapse can be potentiated or depressed depending on the precise relative
timing of the pre- and post-synaptic spikes. 
This timing dependence of magnitude and sign of plasticity, observed in several types of cortical
\cite{markram,feldman,Sjostrom} and hippocampal \cite{biandpoo,Sjostrom,debanne,biandpoo2} neurons, is usually termed Spike
Timing Dependent Plasticity (STDP). 

The role of STDP has been investigated  both in supervised learning framework \cite{Maass},
in  unsupervised framework in which repeating patterns are detected by downstream neurons \cite{masquelier},
cortical development \cite{Song2001},generation of sequences
\cite{newFIETE,newSERGIO} and polychronous activity \cite{newPOLY},
and in an associative memory framework with binary units
\cite{epl2011,WIRN2011}.
However, this is the first time that this learning rule has been used
to make a IF network to work
as associative memory for phase-coded patterns of spike, each of which becomes a dynamic attractor of the network.
Notably, in a phase coded pattern not only the order of activation matters,
but the precise spike timing intervals between units.

We therefore present  a possibility to build a circuit with
 stable phase relationships between the spikes of a population of IF neurons, 
in a robust way  with respect to noise and changes of frequency.
The first important result of the paper is the measurement of the storage capacity of the model, 
i.e. the maximum number of distinct spatiotemporal patterns that can be stored and selectively retrieved,
since it has never been computed in a  spiking model for spatiotemporal patterns. 

Several  classic 
papers (see \cite{SZJ} and references therein) have focused on storage capacity of   binary model with static 
 binary patterns \cite{hopfield}, and much efforts have been done to use more biophysical models and patterns 
\cite{gerstner,SRM,Hopfield95,AlmeidaLisman2007,trevespnas1989,treves2005IF,borisyuk,leibold2006,timme,torcini,NC,epl2011},
but, up to our knowledge,  without any calculation of the storage capacity of spatiotemporal patterns in IF spiking models.
Notably, by introducing an order-parameter which measures the overlap between phase coded spike trains, we are able  
 quantitatively measure of the overlap between the stored pattern and 
the replay activity, and to compute the storage capacity as a function of the model parameters. \\

Another important result is the study of the
different regimes observed by changing the excitability parameters of the network.
In particular, we find that near the region of the parameter space where the network tends to become unresponsive and silent
there is a regime in which the network responds selectively to cue presentation with a short transient replay
of the phase-coded pattern. Differently, in the region of higher excitability,  the patterns
are replayed persistently and selectively, and eventually with more then one spike per cycle.

The paper is organized as follows: Section \ref{sec-model}
introduces the Leaky-Integrate-and-Fire (IF) neuronal model; Section \ref{sec-connections} describes the STDP learning rule used to design the connections; 
in Section \ref{sec-dyn} we study the emergence of collective dynamics and  
 introduce an order parameter to measure the overlap between the collective dynamics and the stored phase coded patterns;
Section \ref{sec-capacity} reports on the storage capacity of the network, i.e. the maximum number
of patterns that can be stored and selectively retrieved in the network; 
the parameter space and the different working regions are  also investigated in Section \ref{sec-capacity};
in Section \ref{sec-noise} we study the robustness of the retrieval dynamics wrt noise and heterogeneity;
Section \ref{sec-oscillators} reports on the implication of this model in the framework of oscillatory interference model
of path-integration; summary and discussion are outlined in Section \ref{sec-summary}.

\section{The model}
\label{sec-model}
We consider a recurrent neural network with $N(N-1)$
possible connections $J_{ij}$, where $N$ is the number of neural units. The connections $J_{ij}$ are 
designed during the learning mode, when 
the connections change their efficacy according to a learning rule inspired to the STDP. 
After the learning stage, the connections values are frozen, and the collective dynamics
is studied.
 This distinction in two stages, 
plastic connection in the learning mode  and frozen connections in the dynamics mode, is a useful framework  to simplify the analysis.
It also finds some neurophysiological motivations 
in the effects of neuromodulators, such as dopamine and acetylcholine \cite{Hasselmo,Hasselmo2}, which regulate 
excitability and plasticity. 

The single neuron model is a Leaky Integrate-and-Fire (IF) \cite{bookG}.
This simple choice, with few parameters for each neuron, is suitable to study the emergence of collective dynamics and
the diverse regimes of the dynamics, instead of focusing on the complexity of the neuronal internal structure.
We use the Spike Response Model (SRM) formulation \cite{bookG,SRM} of the IF model,
which allows us to use an event-driven programming and makes the numerical simulations faster with respect to a differential equation formulation.

In this picture, the postsynaptic membrane potential is given by:
\begin{equation}
h_i(t)=\sum_{j} J_{ij} \sum_{ {\hat t}_j > {\hat t}_i}  \epsilon(t-{\hat t}_j),
\label{IF}
\end{equation}
where $J_{ij}$ are the synaptic connections, $\epsilon(t)$ describes the response kernel to incoming spikes on neuron $i$, and the sum over ${\hat t}_j$ runs over all presynaptic firing times following the last spike of neuron $i$. 
Namely, each presynaptic spike $j$, with arrival time $\hat{t}_j$, is supposed to add to the membrane potential a postsynaptic potential of the form $J_{ij} \epsilon(t-{\hat t}_j)$, where
\begin{equation}
\epsilon(t-{\hat t}_j)= K \left[
\exp\left(-\frac{t-{\hat t}_j}{\tau_m}\right) - 
\exp\left(-\frac{t-{\hat t}_j}{\tau_s}\right) 
\right]
 \Theta(t-{\hat t}_j)
\label{tre}
\end{equation}
where $\tau_m$ is the membrane time constant (here 10 ms), $\tau_s$ is the synapse time
constant (here 5 ms), $\Theta$ is the Heaviside step function,
and K is a multiplicative constant chosen so that the maximum value
of the kernel is 1. The sign of the synaptic connection $J_{ij}$ sets the sign of the postsynaptic potential's change,
so there's inhibition for negative $J_{ij}$ and excitation for positive $J_{ij}$.
 
When the membrane potential $h_i(t)$ exceeds the spiking threshold $\theta_{th}^i$, a spike is scheduled,
and the membrane potential is reset to the resting value zero. 
We use the same threshold $\theta_{th}$ for all the units, except in sec \ref{sec-noise} where different values
$\theta_{th}^i$ are used and the robustness w.r.t. the heterogeneity is studied.
Clearly the spiking threshold $\theta_{th}$ of the neurons is related to the excitability of the network,
an increase of the value of $\theta_{th} $ is also equivalent to a decrease of K, the size of the unitary postsynaptic potential, or, equivalently  to a global decrease in the scaling factor of synaptic connections $J_{ij}$. \\
Numerical simulations of this  dynamics are performed for a network with $P$ stored patterns, 
where connections $J_{ij}$ are determined via a learning rule described in the next paragraph. 
We found that a few number of spikes, given a in proper time order, are able to selectively induce  
the emergence of a persistent collective spatiotemporal pattern, which replays one of the stored pattern (see sec \ref{sec-dyn}).

\section{Designing the connections of the network}
\label{sec-connections}
In a learning model previously introduced in \cite{SZJ,NC,PREYoshioka}, the
average change in the
connection $J_{ij}$, occurring in the time interval $[-t_{learn},0]$ due to  periodic spike trains
of period T, with $t_{learn}>>T$, was formulated as follows:
\begin{equation}
\delta J_{ij} = \frac{T}{t_{learn}} \int\limits_{-t_{learn}}^{0}d t
\int\limits_{-t_{learn}}^{0}d t^\prime \, x_i(t) A(t-t^\prime)
x_j(t^\prime) \label{lr}
\end{equation}
where $T/t_{learn}$ is a normalization factor,
$x_j(t)$ is the activity of the pre-synaptic neuron at time t, and
$x_i(t)$ the activity of the post-synaptic one.
It means that the probability for unit $i$ to have a spike in
the interval $(t,t+\Delta t)$ is proportional to $x_i(t)\Delta t$
in the limit $\Delta t\to 0$.
 The learning window A($\tau=t-t'$) is the measure of the strength of synaptic change
when a time delay $\tau$ occurs between pre and post-synaptic activity.
To model the experimental results of STDP in hippocampal neurons, the  learning window
$A(\tau)$ should be an asymmetric function of $\tau$, mainly
positive (LTP) for $\tau>0$ and mainly negative (LTD) for
$\tau<0$.\\
Equation (\ref{lr}) holds for activity pattern $x(t)$ which represents
instantaneous firing rate, and is suitable to use in 
analog rate models \cite{SZJ,NC,PREYoshioka,Hippo,LNCS} 
and  spin network models \cite{frontiers2010,epl2011}. 
Differently, here, being interested 
in spiking neurons, the patterns to be stored are defined as precise periodic sequence of spikes, i.e. spike-phase coded patterns. Namely, activity of the neuron $j$ is a spike train at times $t^\mu_j$, 
\begin{equation}
x_j^\mu(t)=\sum_n \delta(t- (t^\mu_j+ n T^\mu)),
\label{spiketr1}
\end{equation}
where $t^\mu_j+n T^\mu$ is the set of  spikes times of unit j in the pattern $\mu$ with period $T^\mu$, and frequency $\nu^\mu=1/T^\mu$.
Therefore, following Eq.(\ref{lr}), the change in the connections $J_{ij}$
due to the learning  of the pattern $\mu$ when the time duration of the learning process $t_{learn}$
is  longer then a single period $T^\mu$, is simply given  by
\begin{equation}
J_{ij}^\mu=\sum_{n=-\infty}^{\infty}
 A(t^\mu_j-t^\mu_i+ n T^\mu).
\label{lr2}
\end{equation}
The window $A(\tau)$, shown in Fig.\ 1, is given by
\begin{equation}
A(\tau) = \left \{ \begin{array}{ll}
a_p e^{-\tau/T_p} - a_D e^{-\eta \tau/T_p}\qquad & \textrm{for}\quad \tau>0 \\
a_p e^{\eta\tau/T_D} - a_D e^{\tau/T_D} & \textrm{for}\quad  \tau<0,
\end{array}\right.
\label{At}
\end{equation}
with the same parameters used in \cite{Abarbanel} to fit the experimental data of \cite{biandpoo}, namely
$a_p = \gamma\,[1/T_p + \eta/T_D]^{-1}$,
$a_D = \gamma\,[\eta/T_p + 1/T_D]^{-1}$,
with $T_p=10.2$ ms, $T_D=28.6$ ms, $\eta=4$, $\gamma=0.42$.
This function  satisfies the balance condition $\int_{-\infty}^\infty A(\tau) d\tau =0$.
Notably, when $A(\tau)$ is used in eq. \ref{lr2} to learn phase-coded patterns
with uniformly distributed phases, then the property  $\int A(\tau) dt = 0$
assures that in the connection matrix the summed excitation $(1/N)\sum_{i ,\\J_{ij}>0} J_ {ij}$
and the summed inhibition  $(1/N)\sum_{i ,\\J_{ij}<0} J_ {ij}$
are equal in the thermodynamic limit, and therefore it assures a balance between excitation and inhibition.

Writing Eq.\ (\ref{lr}-\ref{lr2}), implicitly we have assumed that, with periodic phase-coded spike trains
 used to induce plasticity, the effects of all separate spike pairs  sum linearly, each weighted by the same 
STDP window reported in Fig. 1. Timing-dependent learning curves as the one reported in Fig. 1 are indeed typically measured  by
giving an order of 100 pairs of spikes repeatly, with fixed phase relationship, and fixed frequency in a proper range.

However, in different situations, for instance if the frequency is too low or to high \cite{Sjostrom},
or in case of few spike pairs \cite{wittwang2006}, the timing dependence of plasticity is not well 
described by the bidirectional window used here, and a more detailed model is needed to account for 
integration of spike pairs when arbitrary trains are used (see \cite{wangshouval2010,brunel} and references therein).

\begin{figure}[!htbp]
\centering
\includegraphics[width=5cm]{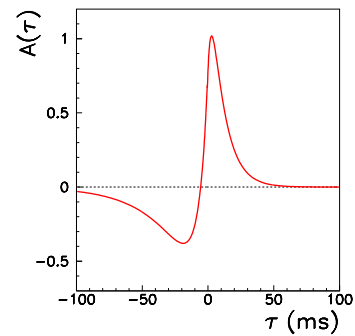}
\caption{a) Plot of the learning window $A(\tau)$ used in the
learning rule (see Eqs. (3), (5), (6)) to model STDP effects. The parameters of the function $A(\tau)$ (Eq. (6)) are 
determined by fitting the experimental data reported in \cite{biandpoo}.}
\end{figure}

The spikes  patterns used in this work are periodic spatiotemporal sequences, made up of 
 one spike per cycle and each of which has a phase $\phi^\mu_j$  randomly chosen  from a uniform distribution in $[0,2\pi)$. 
In each pattern, information is coded in the precise time delay between spikes of unit $i$ and unit $j$, which corresponds to a precise phase
relationship among  units  $i$ and $j$. A spatiotemporal pattern represented in this way is often called phase coded pattern.
Pattern's information is coded in the spiking phases which, in turn, shape the synaptic connectivity responsible of the emerging dynamics and 
the memory formation. \\
The set of timing of spikes of unit $j$  can be defined as $\quad t^\mu_j + n T^\mu =(\phi^\mu_j)/(2\pi \nu^\mu)
+ n/\nu^\mu$, where  
$\nu^\mu$ is the oscillation frequency of the neurons. Thus, each pattern $\mu$ is represented through the 
frequency $\nu^\mu$ and the specific phases of spike $\phi^\mu_j$ of the neurons $j=1,..,N$.
The change in the connection $J_{ij}$ provided by the learning of pattern $\mu$ is given by
\begin{equation}
J_{ij}^\mu=\sum_{n} A(t^\mu_j-t^\mu_i+n T^\mu)=
\sum_n A(\frac{\phi^\mu_j}{ 2\pi \nu^\mu}-\frac{\phi^\mu_i}{2\pi\nu^\mu} +n/\nu^\mu ).
\label{conn}
\end{equation}
When multiple phase coded patterns are stored, the learned
connections are simply the sum of the contributions from individual patterns,
namely  
\begin{equation}
J_{ij}=\sum_{\mu=1}^P J^{\mu}_{ij}.
\label{connP}
\end{equation} 
Note that ring-like topology with strong unidirectional connections  is formed
 only in the case P=1, when a single pattern is stored.  When multiple patterns
are stored in the same connectivity, with phases of one pattern
uncorrelated with the others, bidirectional connections are possible,
and the more the stored patterns, the less the ring-like is the connectivity. 
Even in the cases when the connectivity  is not ring-like  
 the network is still able to retrieve each of the P stored patterns 
in a proper range of threshold values (see storage capacity in Sec \ref{sec-capacity}).

\section{Emerging of collective patterns in the neural dynamics of the network}
\label{sec-dyn}
We study a recurrent network with $N$ leaky Integrate and Fire units, 
 with connections fixed to the values calculated 
in Eqs. (7,8) for different values of P. The results show that, within a well specified range of parameters, our IF 
network is able to work as an associative memory for spike-phase patterns.

In order to check if the network is able to retrieve selectively each of the stored patterns,
we give an initial signal, made up of $M\ll N$ spikes, taken from the stored pattern
$\mu$, and we check if this initial short cue is able to selectively trigger a 
collective sustained activity that is the replay of the same stored pattern 
$\mu$, i.e. checking if the sustained activity has
spikes aligned to the phases $\phi_i^\mu$ of pattern $\mu$.

An example of successful selective retrieval process is shown in Fig. 2 where, depending 
on the partial cue presented to the network, a different collective activity emerges with the 
phases of the firing neurons which resemble one or another of the stored patterns.\\
In this work the cue is a stimulation with  $M$ spikes, with $M=N/10$,
at  times $t^\mu_i= T_{stim} \phi_i^\mu$, $0<i<M$, with $T_{stim}=50$ms. 
In the example shown in Fig. 2 the short stimulation (which lasts less then 5 ms, 
shown in pink in all the figures) has the effect to selectively trigger  the
sustained replay of pattern $\mu$.\\
Note that the retrieval dynamics has the same phase relationship among units than the stored pattern, but
the replay may happen on a  time scale different from the scale used to store the pattern, 
and the collective spontaneous dynamics is a time compressed (or dilated) replay of
the stored pattern. 
Indeed, the period of the collective periodic pattern which emerges during retrieval stage
 may be different then the period of the periodic pattern used in the learning stage.
In the example of Fig. 2 the time scale of
the retrieval dynamics (Fig. 2c,d) is faster then the time scale used to learn the patterns (Fig. 2a,b). 
In the following  we will study the factors affecting the time scale during retrieval, given  
the time scale of the pattern used during learning.
\\
Clearly,  regions of the parameter space in which the network is unable to retrieve selectively the patterns also exist. 
In these regions the retrieval dynamics may correspond to  a mixture of patterns or a spurious state, i.e. a state which is not correlated with any of the stored patterns because the number of stored patterns exceeded the storage capacity of the network.
As discussed below, the storage capacity, defined as the maximum number of encoded   and successfully retrieved patterns, 
depends on the frequency used during the learning stage (which affects connectivity), and on the spiking threshold of 
the units (which affects excitability and network dynamics).

Example of failure are shown in Fig.3.
In Fig. 3b the network has too low excitability and the response is not persistent, while
in Fig. 3a the emerging dynamics is not correlated with any of the stored patterns. 
\begin{figure}[t!]
\begin{center}
\includegraphics[width=4.2cm]{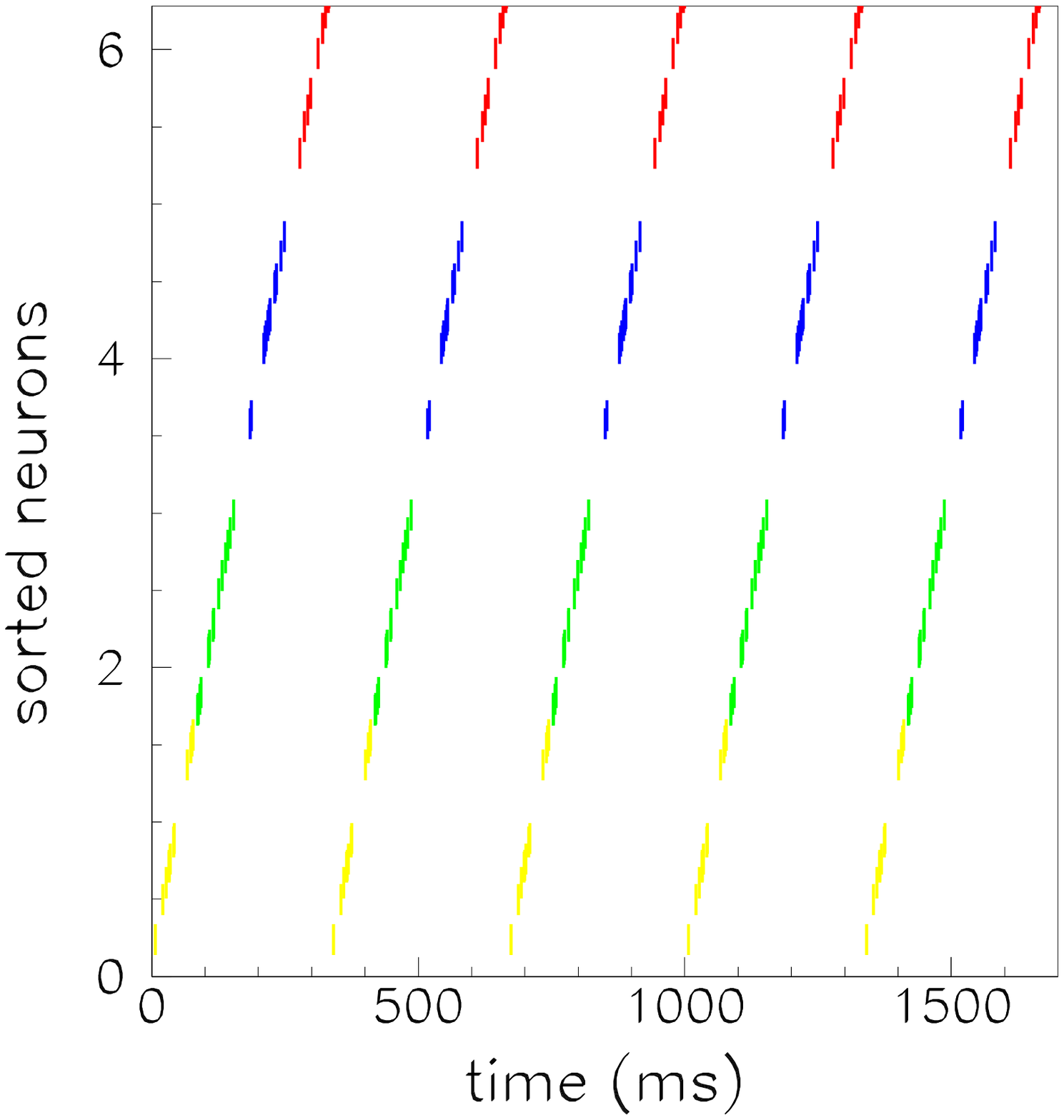}
\includegraphics[width=4.2cm]{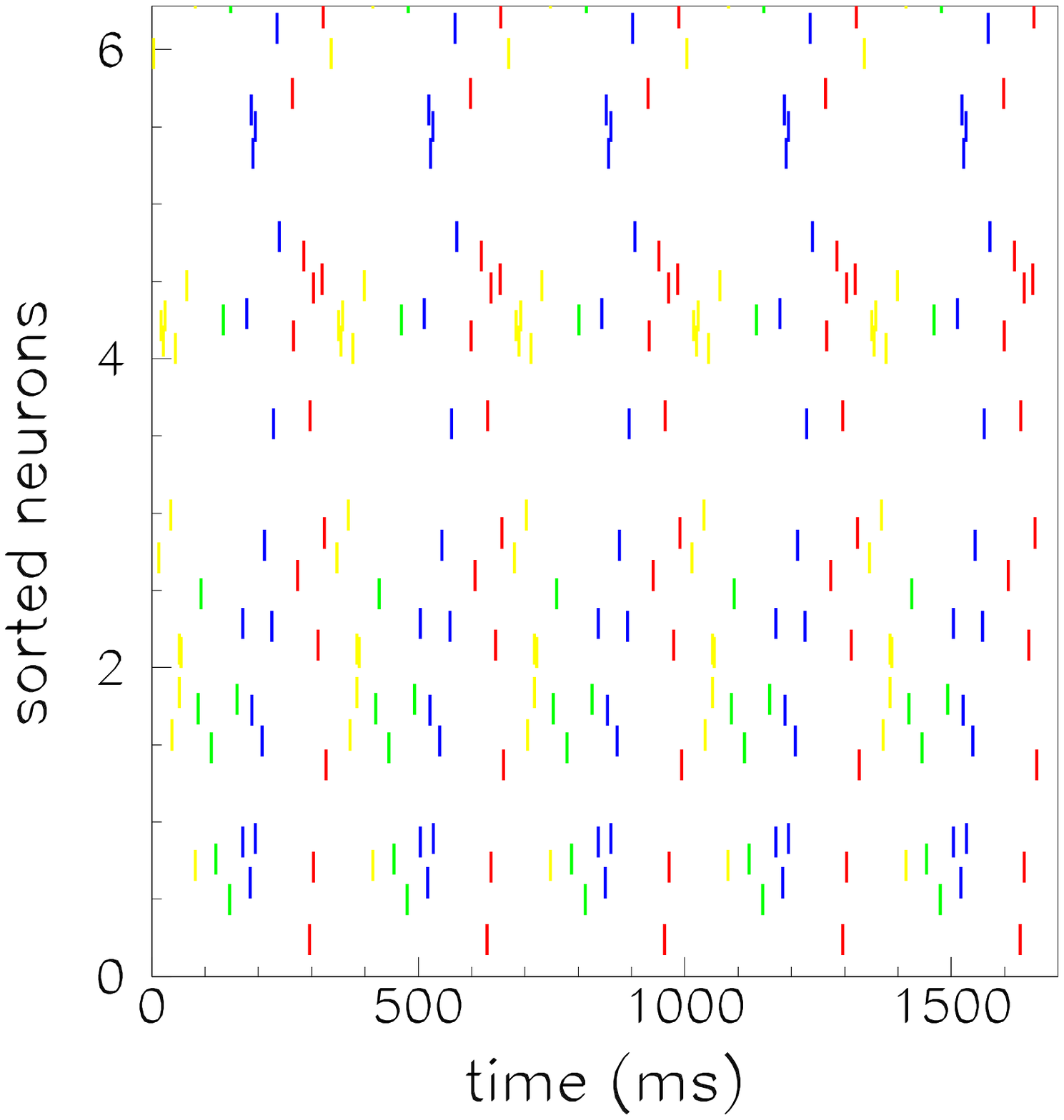}
\\
a) \quad \quad \quad \quad \quad \quad \quad \quad \quad \quad  b)\\
\includegraphics[width=4.2cm]{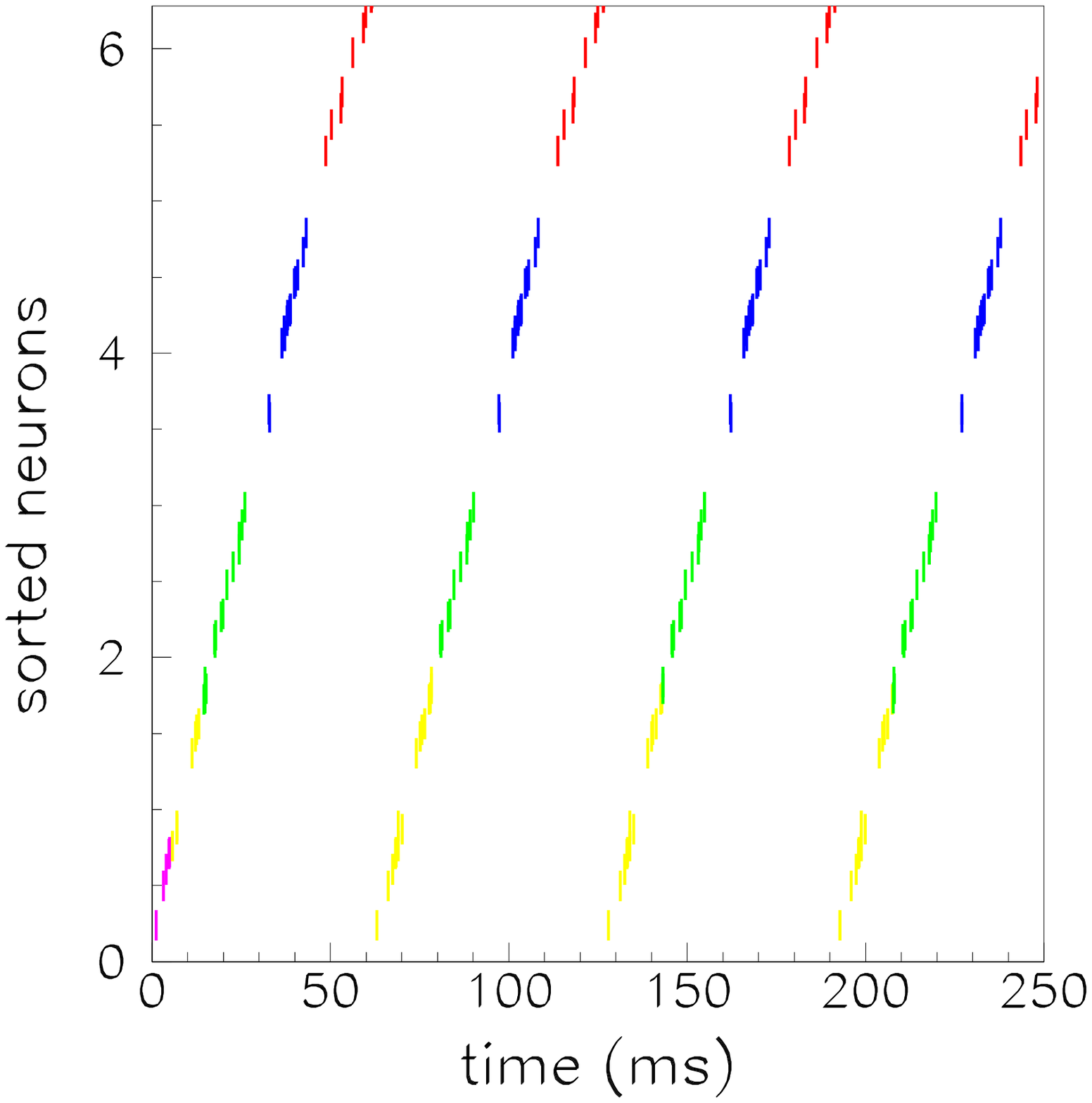}
\includegraphics[width=4.2cm]{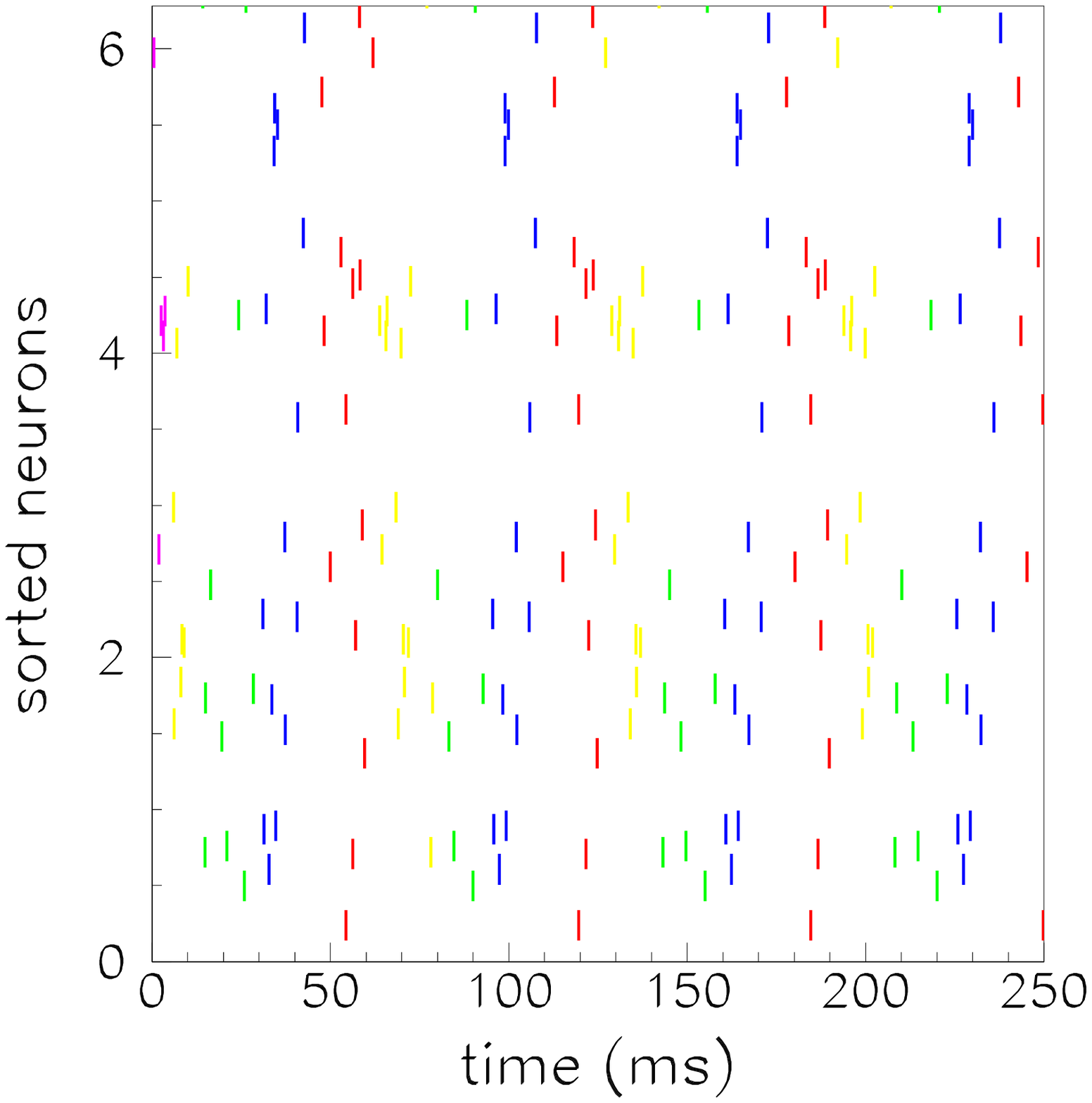}
\\
c) \quad \quad \quad \quad \quad \quad \quad \quad \quad \quad  d)
\end{center}
\caption{
Examples of selective successful retrieval (c,d) of two stored patterns (a,b).
The raster plot of  50  units (randomly chosen) are shown sorted on the vertical axis 
according to increasing values of phase $\phi_i^1$ of the first stored pattern $\mu=1$.
The network has $N=3000$ IF neurons, $\Theta_{th}= 70$ and
connections given by Eq. (7,8) with $P=5$ stored patterns at $\nu^\mu= 3$ Hz.
Two of the stored patterns used during the learning mode are shown in a,b.
The dynamics emerging after a short train of
$M=N/10$ spikes with phases similar to the pattern shown in a,b,
is shown in c,d respectively.
The dynamics of the network, after a transient, is periodic of period $T$.
The spikes which belong to the trigger are shown in pink in (c,d),
 the other different colors represent the 
value of $t_i/T \,mod\, 4$, where $t_i$ is the time of the spike of
the unit i during the emerging spontaneous dynamics.
Figure c shows that when
the network dynamic is stimulated by a partial cue of pattern
$\mu=1$, the neurons oscillate with phase alignments resembling
pattern $\mu=1$, but at different frequency. Otherwise, in  d,
when the partial cue is taken from pattern $\mu=2$, the neurons
phase relationships, even if periodic,  are uncorrelated with
pattern $\mu=1$, and recall the phase of pattern $\mu=2$.
}
\label{fig_varphi}
\end{figure}
\begin{figure}[t!]
\begin{center}
\includegraphics[width=4.2cm]{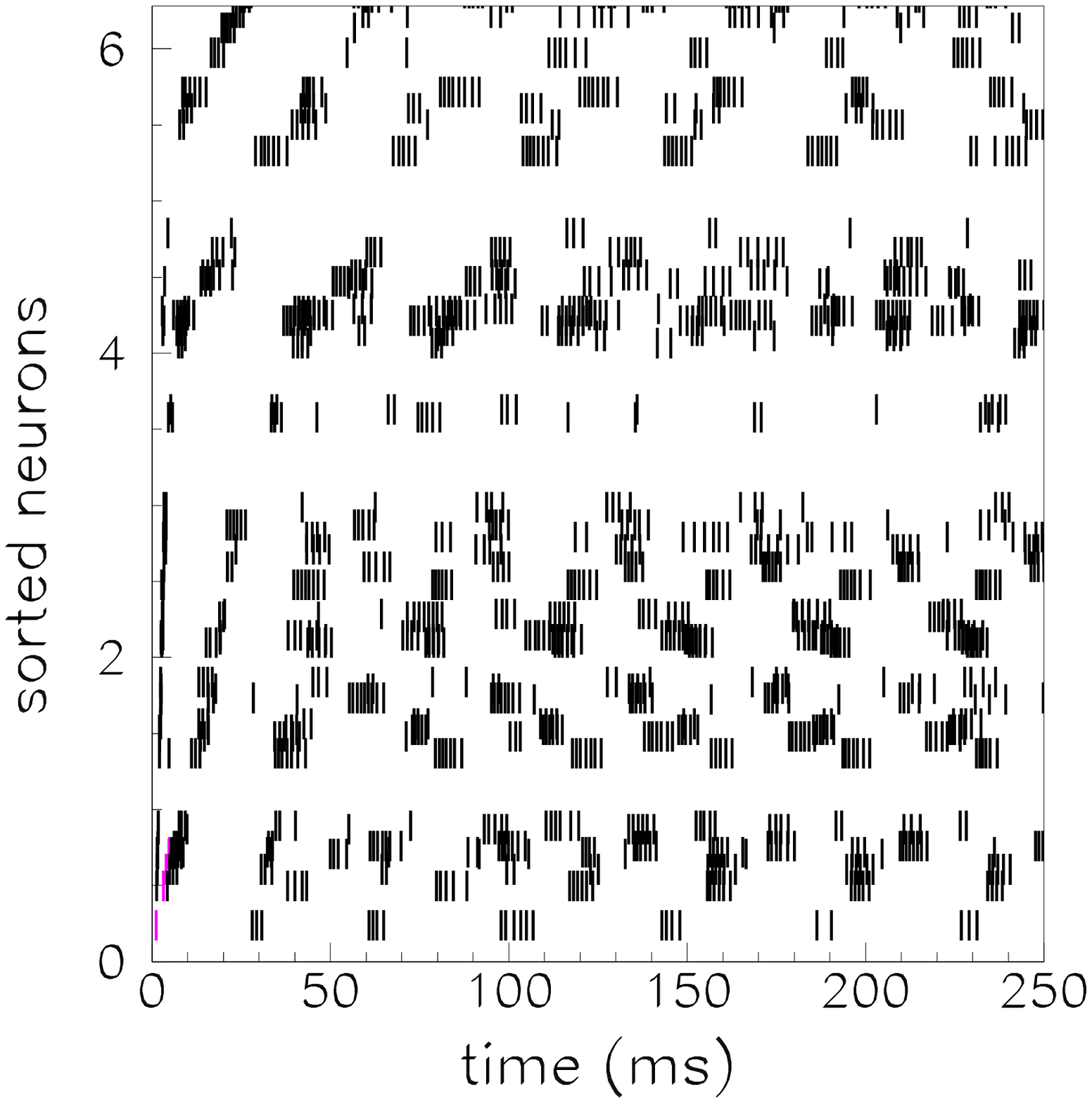}
\includegraphics[width=4.2cm]{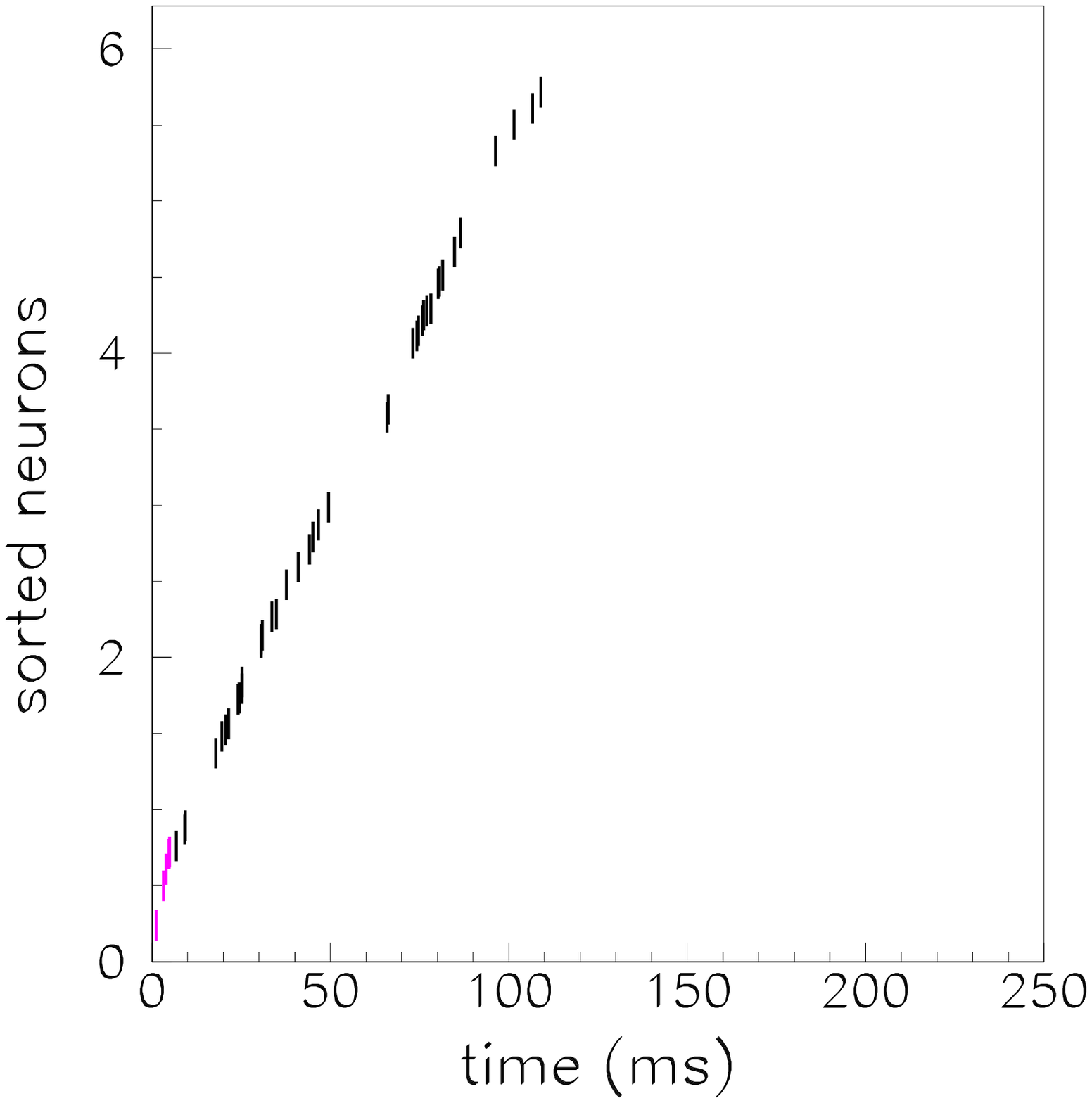}
\\
a) \quad \quad \quad \quad \quad \quad \quad \quad \quad \quad  b)
\end{center}
\caption{
 Example of neural response  in two case of failure of retrieval.
A spurious state emerge in (a), while a short transient response emerges in the case shown in (b).
 $N=3000$ and $\nu^\mu=3 Hz$ as in previous picture,
while the values $\theta_{th}$ and $P$ are 
   $\theta_{th} = 10,P=5$ in (a) and
 $\theta_{th} = 95,P=5$ in (b).
The dynamics emerging after a short train of
$M=N/10$ spikes with phases equal to the stored pattern (pattern shown in Fig 2a), 
is not a self-sustained retrieval of the pattern.
For clarity, the raster plot of only 50 (randomly chosen) units are shown, sorted according to
increasing value of phase ${\phi_i}^1$ of the stored pattern. 
}
\label{fig_failure}
\end{figure}\\
To measure quantitatively the success of the retrieval, in analogy with the Hopfield model,
we introduce an order parameter, which estimates the overlap  between the network collective activity
during the spontaneous dynamics and the stored phase-coded pattern.
This quantity  is $1$ when the phases $\phi_j$ of neurons $j$ coincides
with the stored phases $\phi_j^\mu$, and is close to zero when the phases are uncorrelated with the stored ones.
Therefore, we consider the following dot product $|m^\mu(t)|=<\xi(t)|\xi^\mu>$ where $\xi^\mu$ is
the vector having components $ e^{i \phi_j^\mu}$, namely: 
\begin{equation}
|m^\mu(t)|= \left|\frac{1}{N}\sum_{\stackrel{j=1,\ldots,N}{\scriptscriptstyle t-T^*<t_j^*<t}}  e^{-i 2 \pi t_j^*/T^*}  e^{i \phi_j^\mu}\right|
\label{nn}
\end{equation}
where $t_j^*$ is the spike timing of neuron $j$ during the
spontaneous dynamics, and $T^{*}$ is an estimation of the period of
the collective spontaneous periodic dynamics. 
 The overlap in Eq. (\ref{nn})
is equal to $1$ when the phase-coded pattern is perfectly retrieved
 (i.e. same sequence and phase relationships among spikes,
 even though on a different time scale), while is of order
$\simeq 1/\sqrt{N}$  when phases of spikes are uncorrelated to the
stored phases. The  order parameter $|m^{\mu}|$ allow us to measure 
the network storage capacity in the space of parameters $\theta_{th}$ and $\nu^{\mu}$.\\
Note that the value of  $m^\mu(t)$ between two periodic spike trains measures
the similarity in the sequence of spiking neurons and in the
 phase lag between the spikes, being invariant by a simple change in time scale.
This is a suitable choice especially when the replay of a spatio temporal
pattern has to be detected independently from the compression of the time scale.
Note that if we have a spike train  that is not periodic, we cannot define the
period,  however we can define the order parameter (\ref{nn}) 
looking at the time-window $T^*$ which maximize the order parameter. This
can be useful in the case when one looks for a short replay hidden in a  not-periodic spike train,
 such in many experimental situations.  \\
The value of $m^\mu(t)$ after a transient converges to a stable value which 
is  close to one when pattern $\mu$ is retrieved (for example in Fig. 2c at large times $m^{\mu=1}=1$, and $m^{\mu=2}=0.01$)
while $m^\mu(t)$  is of order $\simeq 1/\sqrt{N}$ for all $\mu$ in the case of failure of retrieval.
Two further cases of failure can occur:  in  Fig. 3a $m^\mu(t)$ after the transient has values in the range $0.01 - 0.02$ for all $\mu$ because
the emerging dynamics is a spurious state not correlated with any of the stored phase-patterns, while $m^\mu(t)$ is zero in Fig. 3b  
since the network becomes silent.\\
In the following, the storage capacity of the model is analyzed  considering
the maximum number of patterns that the network is able to store and selectively recall. 
In particular we investigate the role of two  model parameters: the frequency of the stored patterns $\nu_\mu$,  
and the spiking threshold $\theta_{th}$ affecting the excitability of the network. \\
\section{Storage Capacity}
\label{sec-capacity}
Numerical simulations of the  IF network  with $N=3000$ neurons  were performed by 
systematically changing the value of the spiking threshold $\theta_{th}$, the connections $J_{ ij}$, and for different number of patterns P and frequency $\nu_{\mu}$. Here we propose results for a unique value of the spiking threshold $\theta_{th}$ for all neurons,
however the behavior is also robust with respect to a variability in the threshold values among neurons, as reported in the next Section.\\
Network storage capacity is defined as $\alpha_c = P_{max}/N$, where N is the number of neurons and $P_{max}$ is the maximum number of patterns that can be stored and successfully retrieved with an overlap $|m^\mu|$ larger than a certain value, which measures the degree of similarity. 
Given that in our simulations the overlap $|m^\mu(t)|$ at large times has mostly two possible values, close to one (success) or close to $1/\sqrt N$ (failure),
we  fixed the desired similarity  value to 0.5, since the whole storage capacity analysis is very robust with respect to  this
parameter (since the transition between low values and high values of $|m^\mu(t)|$ as
a function of P is sharp). \\ 
Patterns with random phases were  extracted and used to   define the network connections $J_{ij}$ with the rule Eq. (\ref{connP}). 
After the stimulation with a short train of $M=N/10$ spikes taken at times $t_i$ from the first pattern,  the dynamics is simulated and the overlap defined in Eq. (\ref{nn}) with $\mu = 1$ is evaluated at large times. 
If the  overlap $|m^{\mu=1}(t)|$, averaged over 50 runs, is greater than $0.5$ at time $t>\bar t$  (where $\bar t= 600$ ms for all the simulations), then we consider the retrieval  successful for that pattern.  The maximum value of P, for which the network is able 
to successfully replay  each of the stored patterns, defines the storage capacity of the network.\\
\begin{figure}[ht]
\begin{center}
a)\quad
\includegraphics[width=6.5cm]{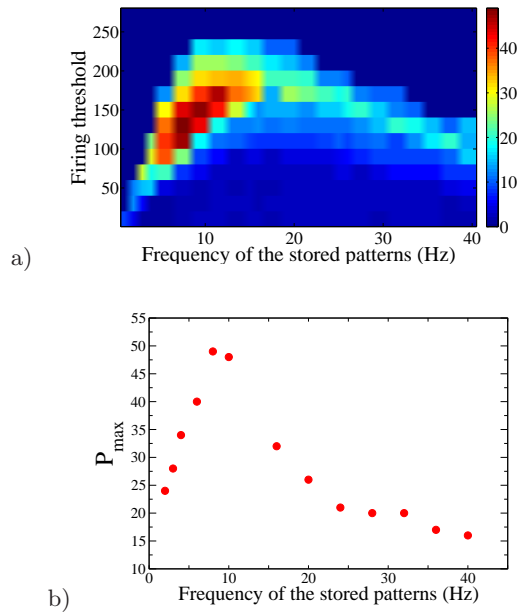}\\
\vspace{0.5cm}
b)\quad
\includegraphics[width=5.5cm]{storage_varf6.eps}
\end{center}
\caption{(a) Storage capacity in a network of $N=3000$ units, as a function of the
spiking threshold $\theta_{th}$ and oscillation frequency $\nu^\mu$ of stored patterns.
 The maximum number of  patterns 
successful retrievable $P_{max}$ is shown in color-coded legend,  
the value grows from $P_{max}=0$ (dark blue) to $P_{max}=50$ (strong red).
(b) The storage capacity $P_{max}$
 as a function of the
frequency of stored patterns, once fixed the threshold $\theta_{th}$ to the optimal value for each frequency. 
} 
\label{Fig7}
\end{figure}
\begin{figure}[ht]
\begin{center}
a)\includegraphics[width=6.5cm]{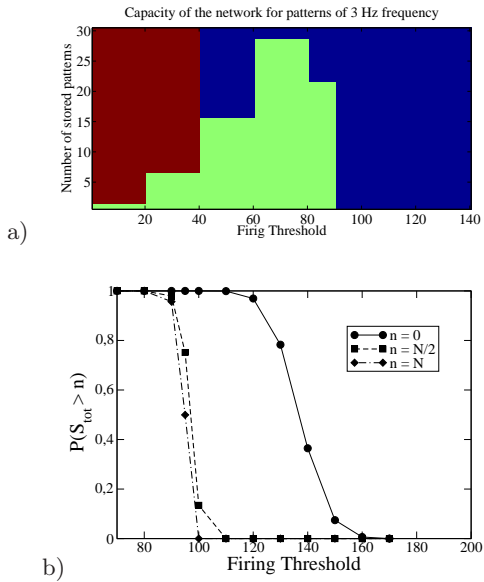}\\
\vspace{0.5cm}
b)\,\includegraphics[width=5.5cm]{prob_2.eps}
\end{center}
\caption{a) Storage capacity at $\nu^\mu= 3$ Hz:  
 the region of successful retrieval  as a function of spiking threshold and number of patterns is marked in green.
The region with persistent activity not correlated with any of the stored pattern is marked in red (spurious states), and the region
in which the network responds with only a short transient and then becomes silent is marked in blue (see examples in fig 3).
b) The probability that the size $S_{tot}$ of the network response,
 measured as the number of the spikes that follow the cue stimulation,
is larger than $n$, with $n=0,N/2,N$, is shown as a function of spiking threshold $\theta_{th}$, in a network with $\nu^\mu=3 Hz$
and $P=1$.
As always in this paper the number of units is $N=3000$. 
The figure shows that near $\theta_{th}^{crit} \simeq 90$  there's a transition from a region of persistent replay to a region 
of silence.
 }
\label{Fig6}
\end{figure}
The storage capacity as a function of the spiking threshold $\theta_{th}$ 
and storing frequency $\nu^\mu$ 
is reported in Fig. \ref{Fig7}a, where $P_{max}$ is shown in a color-coded legend.
The largest capacity is achieved when the frequency of the stored patterns during learning is $\nu^\mu \simeq 8 $ Hz
 and the spiking threshold of the units during retrieval is $\theta_{th} \simeq 130$, which provides a capacity $\alpha_{max}=P_{max}/N=0.016$.
\\
In Fig. \ref{Fig7}b we show 
the storage capacity $P_{max}$ 
as a function of the frequency
$\nu_{\mu}$ once fixed the value of the threshold $\theta_{th}$ 
to the optimal value, corresponding to highest capacity for each frequency.
\\
The optimal storing frequencies and threshold values depend on the time constants of the model, such as the $\tau_s,\tau_m$ of the IF units and the temporal shape of the learning kernel $A(\tau)$, whereas different shapes of $A(\tau)$ may subserve to different storing frequency ranges.
In this work $\tau_s,\tau_m$ and $A(\tau)$ are set to the values described in Sec. $\ref{sec-model}$,
 and the emergent collective dynamics is studied  as a function of the other network parameters.
Indeed, Fig. \ref{Fig7}b shows that for the learning kernel $A(\tau)$ used here, 
there is  peak in the storage capacity around $8$ Hz,
in the range $2$Hz-$20$Hz.
Figure \ref{Fig7}a  also proves that, for each stored frequency, 
a large interval of spiking threshold values $\theta_{th}$ 
exists for which the network is still able to work properly as associative memory for phase-coded patterns.
\\
The associative memory properties as a function of the spiking threshold are reported in Fig. \ref{Fig6}, when
the oscillation frequency of the patterns stored during learning is $\nu^\mu=3$Hz.
The region marked in green in Fig. \ref{Fig6} corresponds to  cases in which the retrieval is successful and 
the cue is able to selectively
activate the self-sustained replay of the stored pattern (with an order parameter $m^\mu$ larger
than $0.5$).
When spiking threshold changes in the range $10<\theta_{th}<90$ the storage capacity
changes between $P_{max}/N=1/3000$ and $P_{max}/N=29/3000$.
\\
Outside the green region  the number of patterns exceeds the  
storage capacity and the retrieval fails.
There are two possible reasons for this behavior.
At low threshold, when the number $P$ of patterns exceeds the storage capacity $P_{max}$,
the network responds with a self-sustained activity that is not correlated with any  of the
stored patterns, i.e. a spurious state.
In this regime, marked with red color in Fig. 5a,
 the order parameter $m^\mu(t)$ is of order $1/\sqrt N$ for all the stored patterns 
(see also raster plot in Fig.3a).
 On the contrary, in the high  $\theta_{th}$ regime, the
network tends to become silent and unresponsive.
Indeed, in the region marked with blue color,
the network responds to the initial cue stimulation with
a short transient and then became silent. In this case the value of $m^\mu(t)$ is zero  because there is no self-sustained 
activity at time $t>\bar t$, meaning that the stored attractors become unstable when $\theta_{th}$ is too high (see raster plot in Fig.3b). \\
For values of the threshold greater than  $\theta_{th}^{crit}$, independently from P,  the network activity 
is never persistent, as reported in Fig. \ref{Fig6}a where $\theta_{th}^{crit}=90$ .\\
At thresholds close to this critical value the network responds with a transient activity that is 
a short replay of the stored pattern, but not a persistent replay. 
The size $S_{tot}$ of the network response, measured as the number of  spikes that follow the cue stimulation, is reported in Fig. \ref{Fig6}b as a
function of $\theta_{th}$, for a network with $\nu^\mu=3$ Hz, $N=3000$ and $P=1$.
\begin{figure}[ht]
\begin{center}
a)\,
\includegraphics[width=3.5cm]{region6.eps}
b)\,
\includegraphics[width=3.5cm]{freqout5.eps}\\
\vspace{0.5cm}
c)\, 
\includegraphics[width=3.5cm]{numbspks3.eps}
d)\, 
\includegraphics[width=3.7cm]{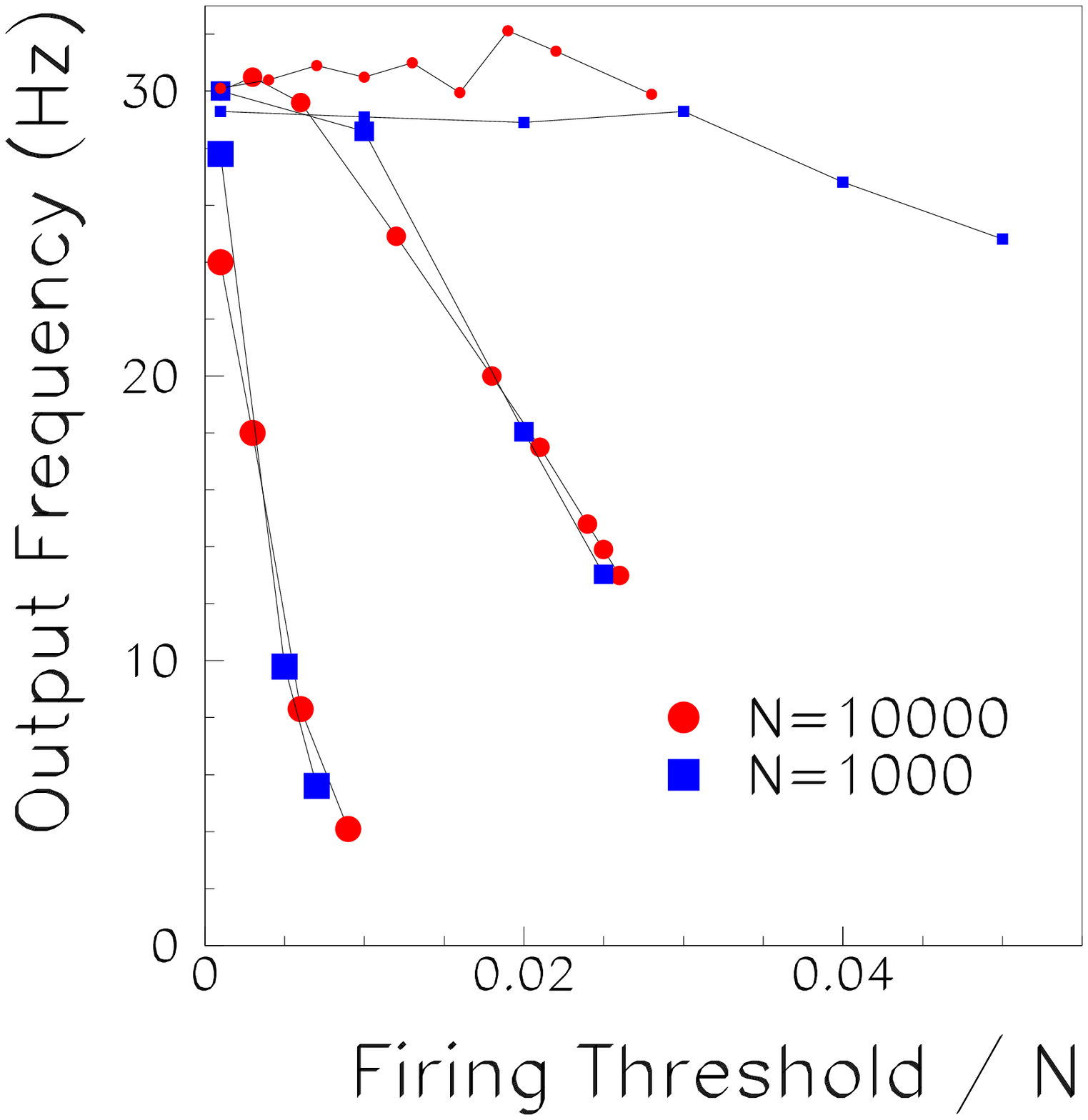}\\
\end{center}
\caption{
(a) Frequency of the collective dynamics during replay 
 as a function of the 
frequency of stored patterns in the network with N=3000 units. 
Dots refer to replay frequency observed at optimal spiking threshold.
The bars refers to the range of frequency available
 through changes in spiking threshold. 
(b)  Frequencies of the collective dynamics during replay
 as a function of the spiking threshold and for different 
stored frequencies (see colors legend). 
Pattern is replayed on a time scale which becomes faster
 if we decrease threshold 
$\theta_{th}$ for the most of the frequencies $\nu^\mu$. 
The dependence is much
stronger for $\nu^\mu \le 4 $ Hz. N=3000.
(c) The number of spikes per cycle as a function of the
spiking threshold $\theta_{th}$ in networks with different frequency 
$\nu^\mu$ of stored pattern.
(d) Frequency of the dynamics during replay as a function of 
ratio between spiking threshold and network size N.
Red dots are results for a network with N=10000 units,
while blue squares are results for a network with N=1000 units.
Size of the symbols refers to the stored frequency,
small symbols (on the top of the picture)
 correspond to stored frequency $\nu^\mu=10 Hz$,
medium size symbols correspond to stored frequency $\nu^\mu=3 Hz$,
and large symbols (bottom) correspond to stored frequency $\nu^\mu=1 Hz$. 
}
\label{Fig-freq}
\end{figure}
\\
In the following we investigate the replay activity  in the region with successful retrieval.
We focused on the dependence of the frequency of collective oscillations during replay on the model parameters.
 Fig. \ref{Fig-freq}a shows the  collective frequency of replay  
as a function of the frequency $\nu^\mu$ of the patterns
 stored in the learning stage with N=3000. 
The red dots in the figure refer to the frequencies of oscillations observed during retrieval at the optimal spiking threshold 
(where the maximum storage capacity occurs), while the bar indicates the available range of frequencies of replay,
 accessible through a change in the spiking threshold. 
Important to note is that, in most of the cases,  the frequency of the stored pattern and the collective replay frequency do not coincide, since the pattern is replayed compressed (or dilated) in time, on a time scale dependent on the network parameters.
We observe that for the chosen parameters $\tau_m, \tau_s$  of the network,
and the given shape of $A(\tau)$,  the replay occurs on a compressed time scale for all stored patterns of frequency lower then $25$ Hz,
while the two time scale coincide when   $\nu^\mu \simeq 25$ Hz.
The dependence of the frequency of the collective oscillations  on the spiking threshold
 is shown in Fig. \ref{Fig-freq}b. This dependence is weak  
 for stored frequencies higher than $10$ Hz.
Besides, for low stored frequencies (1-4 Hz) the
frequency of the replay is very sensitive to the threshold value, 
changing from  $6$ Hz at high spiking threshold to $30$ Hz at low threshold.
\\
We also investigate the frequency's dependence on network size $N$.
In fig. \ref{Fig-freq}.d red dots are results for a network with $N=10000$ units,
while blue squares are results for a network with $N=1000$ units.
If what counts is only the  time lag between  the single units consecutive in the sequence, then one expects that
result with 1Hz stored at $N=10000$ would be similar to 10Hz stored at $N=1000$,
and this is not the case.
We see that when we use a storage frequency equal to 10 Hz
(which corresponds to different time lag between cells depending on N), then
the oscillation frequency during replay is around 30Hz in both networks (both $N=1000$, and $N=10000$),
while, on the other hand, if we have a storage frequency equal to 1 Hz the 
oscillation frequency during replay may span  a large range (5Hz -25Hz) in both networks.
 Fig. \ref{Fig-freq}.d  also shows that frequency of replay
depends on the ratio between spiking threshold  and network size N, and that
 the high sensitivity on spiking threshold value 
holds, when stored frequency is low
(1-4 Hz), also at different values of the network size.
\\
This open the possibility to govern the oscillation frequency of the collective replay activity 
via neuromodulators which change
the excitability and therefore the spiking threshold of the neurons.
Since in our model (see Eq. 1,2) a change of the threshold
is equivalent to a change in the scale factor of all synaptic connections, a similar effect might be achieved also by simply
  driving the cells more due to increased synaptic input.
Importantly, the sensitivity of collective oscillation frequency on spiking threshold is not a sensitivity of the single unit but
of the collective behavior, since, as  discussed in Sec. \ref{sec-noise}, if we change the spiking threshold of few units
the collective rhythm is still unique for the whole population. The replay frequency   depends on the average threshold among units,
but all the units have the same oscillation frequency during replay.\\
Moreover, for networks with $\nu^\mu\ge 10$ Hz, whose replay frequency does not considerably change  with spiking threshold,
the replay dynamics is still affected by the spiking threshold. Indeed, in this case, the number of spikes per cycle
increases with lowering of the spiking threshold.
An example is reported in Fig. \ref{Fig8}. The raster plots show  the same pattern replayed in three  networks having different values of the spiking
 threshold $\theta_{th}$:  a burst of activity takes place within each cycle,
with phases aligned with the pattern, with a number of spikes per cycle dependent on the value of $\theta_{th}$. 
This behavior is summarized in Fig. 6c where the number of spikes per cycle is reported as a function of spiking threshold, at different values of stored frequencies. 
Therefore, by lowering the spiking threshold the replay activity occurs with more than one spike per cycle,
or on a faster time scale (see Fig. 6b,c).\\
\begin{figure}[ht]
\begin{center}
a)\,
\includegraphics[width=3.8cm]{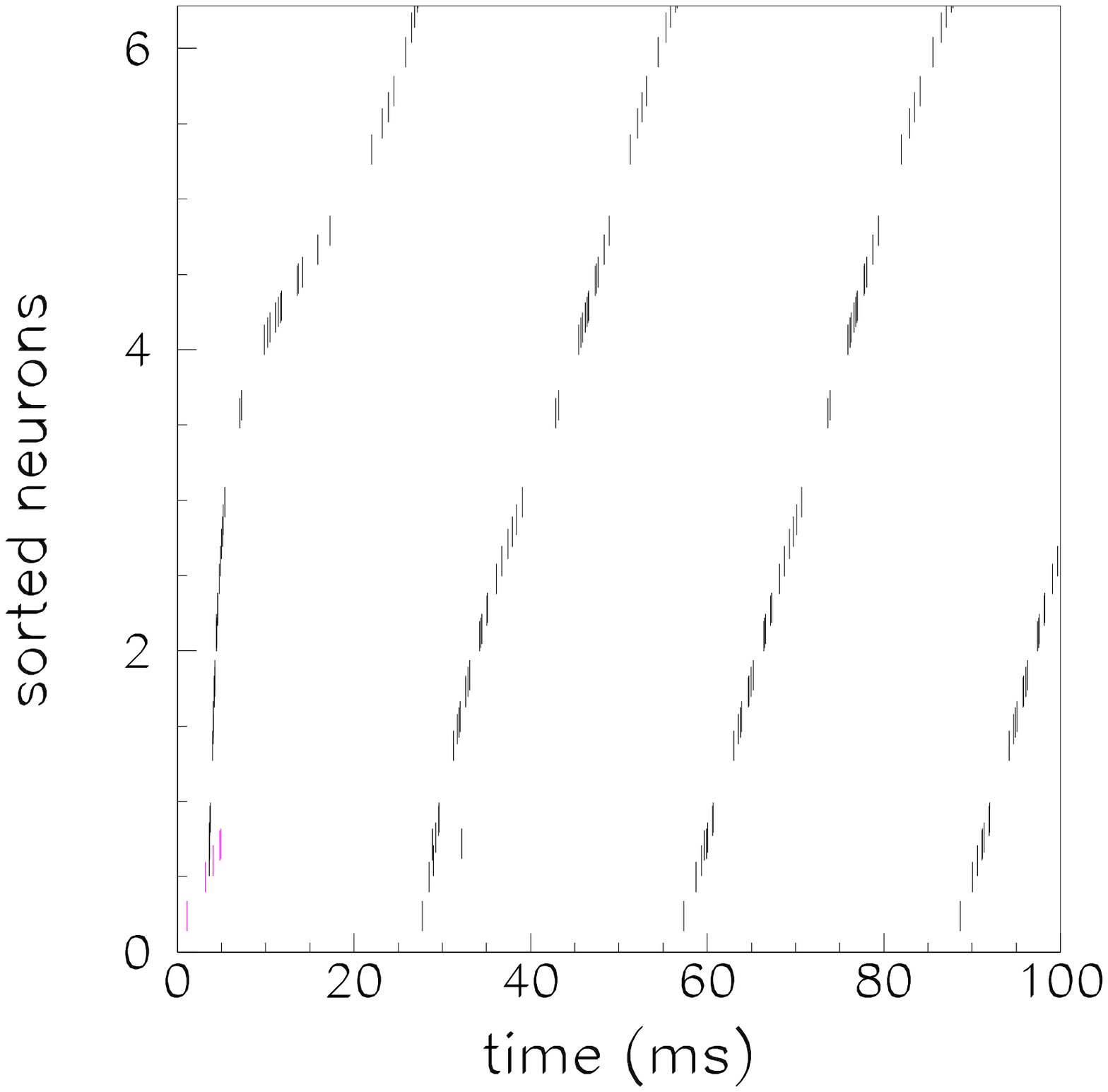}
b)\,
\includegraphics[width=3.8cm]{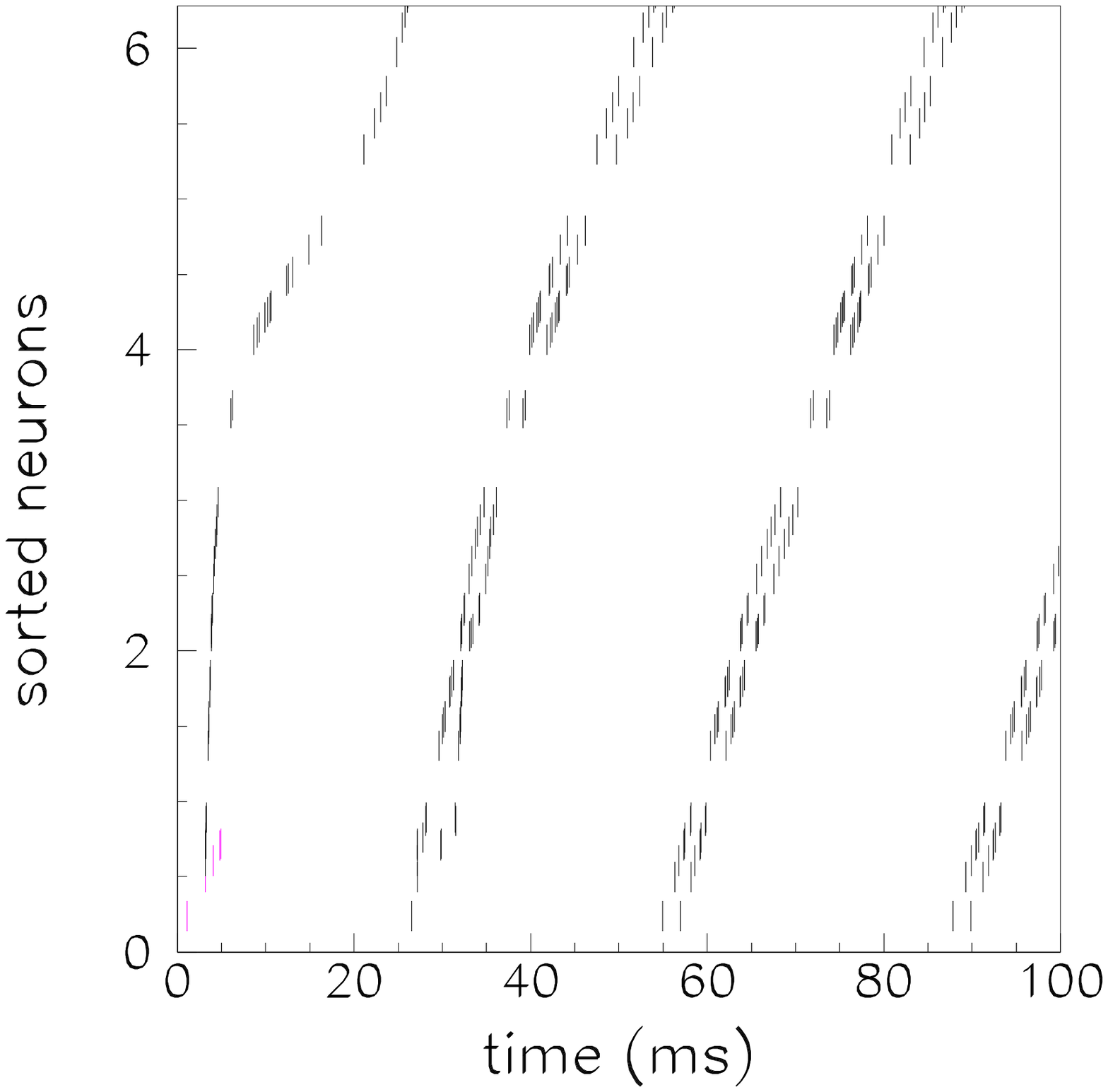}\\
\vspace{0.5cm}
c)\,
\includegraphics[width=4cm]{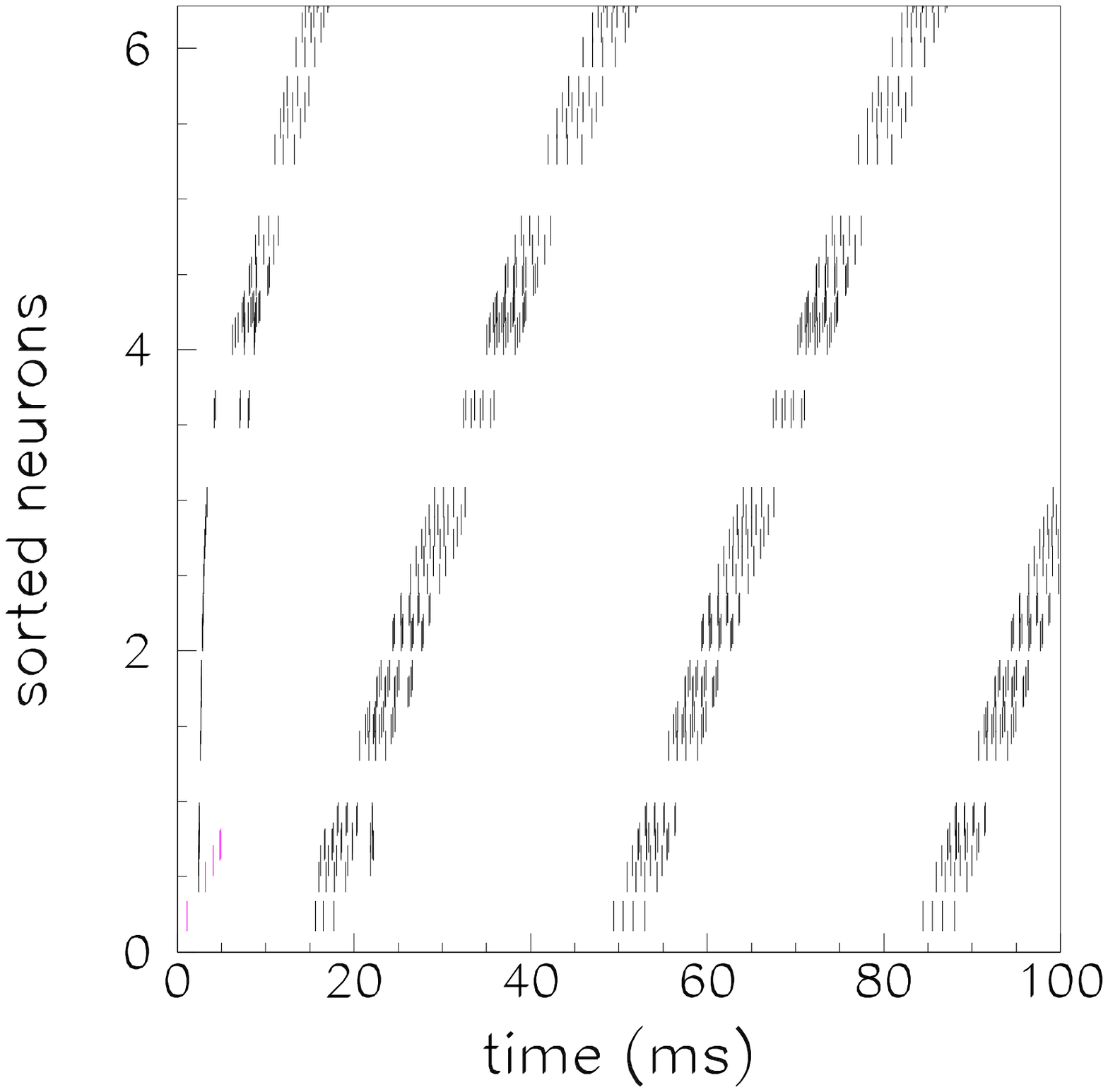}
\end{center}
\caption{%
Modulation of spiking threshold  changes the number of spikes per cycle, 
keeping preserved the phase relationship among units.
Recall of the pattern $\mu=1$ for networks of $N=3000$ units, $\nu^\mu=20$ Hz and different values of
spiking threshold $\theta_{th}=80,65,40$ is shown respectively in a,b,c. Depending on the value of the
 spiking threshold $\theta_{th}$, the phase-coded 
pattern is replayed with a different number of spikes per cycle. 
 Spike of the cue stimulation are shown in pink, while the response of the network in black.
For clarity, the raster plot of only 50 (randomly chosen) units are shown, sorted according to
increasing value of phase ${\phi_i}^1$ of the stored pattern. 
}
\label{Fig8}
\end{figure}
The behavior of the output oscillation frequency suggests that a parameters region exists where the network always 
responds with one spike per cycle. In this region an increase of the  excitability produces a growth of the frequency of oscillation 
up to a plateau value. 
Differently, for higher excitability the frequency does not increase while the number of spikes per cycle grows.
This means that the different frequencies, in addition to the information coded in the phase relationship, can code other information
in relationship with the level of spiking threshold:
at high frequency the threshold changes the number of spikes per cycle, while at low frequency
the threshold changes the frequency of the collective oscillations during the replay.\\
This open the possibility to have a coding scheme in which, while the phases encode pattern's
information, a change in frequency or a change in rate in each cycle 
represents the strength and saliency of the retrieval or it may encode another variable \cite{MATE}. 
The recall of the same phase-coded pattern with a different number of
spikes per cycle is particularly interesting at the light of recent observations of Huxter
{\em et al.} \cite{huxter_burgess} in hippocampal place cells,
showing the occurrence of the same phases with different rates.  The authors prove 
 that the phase of firing and firing rate are dissociable and
can represent two independent variables, e.g. the animal location
within the place field and its speed of movement through the
field.\\  
Notably, the recall of the same phase coded pattern with  different frequencies of oscillation is 
also  relevant and accords well with the need to have stable precise phase relationship among cells
with frequency of oscillation modulated by parameters such as the speed of the animal \cite{Blair2008,Welday2011}. \\

The value of the frequency of collective activity during the replay clearly is
related not only to the threshold and the stored frequency, but also to the 
shape of the learning window $A(\tau)$ and on the two characteristic times
of the model $\tau_s$, $\tau_m$.  A systematic study of the dependence of the replay time scale on the 
shape of STDP and the characteristic times of the neuron model has not yet done in a spiking model,
however a dependence on the asymmetry of $A(t)$ has been analytically found in a simple model
with analog neurons and a single characteristic time \cite{PREYoshioka}. 
 
\section{Effects of noise and robustness of collective oscillation frequency and phase relationships}
\label{sec-noise}
While in the Hopfield model the patterns are static, 
and information is coded in a binary pattern
$\bold S^\mu= S_1^\mu,\ldots,S_N^\mu$, with $S_i^\mu \in \{\pm 1\}$, 
here, in this study, the patterns are  time dependent,
and information is coded in the phase pattern  
$\bold{\phi}^\mu=\phi_1^\mu,...,\phi_N^\mu$ with $\phi_i^\mu \in [0,2\pi]$,
where the value $\phi_i^\mu /(2\pi\nu^\mu)$ represents the time shift of the spike of unit $i$ with respect
to the collective rhythm, i.e the time delay among units.
However, as for the Hopfield model, the patterns  stored in the network are attractors
of the dynamics, when $P$ do not exceeds storage capacity, and the 
dynamics during the retrieval is robust with respect to noise.
We firstly check robustness w.r.t. input noise, i.e  when  a Poissonian noise $\eta_i(t)$
is added to the postsynaptic potential $h_i(t)$ given in Eq.\ref{IF} .
The total postsynaptic potential  of each neuron $i$ is then given by
\begin{equation}
h_i(t)=\eta_i(t)+\sum_{j}J_{ij}\sum_{{\hat t}_j > {\hat t}_i}  \epsilon(t-\hat{t_j})
\end{equation}
where $\eta_i(t)$ is modelled as 
\begin{equation}
\eta_i(t)=J_{\text{noise}}\sum_{\hat{t}_{\text{noise}}>\hat{t}_i} \epsilon(t-\hat{t}_{\text{noise}}).
\end{equation} 
The times $\hat{t}_{\text{noise}}$ are randomly  extracted for each neuron $i$,
and $J_{\text{noise}}$ are random strengths, extracted independently for each neuron $i$ and time $\hat{t}_{\text{noise}}$.
The intervals between times $\hat{t}_{\text{noise}}$ are extracted from a Poissonian
distribution $P(\delta t) \propto e^{-\delta t/(N\tau_{\text{noise}})}$, while
the strength  $J_{\text{noise}}$ is extracted from a Gaussian distribution with mean ${\bar J}_{\text{noise}}$
and standard deviation $\sigma(J_{\text{noise}})$.

\begin{figure}[t!]
\begin{center}
\includegraphics[width=9cm]{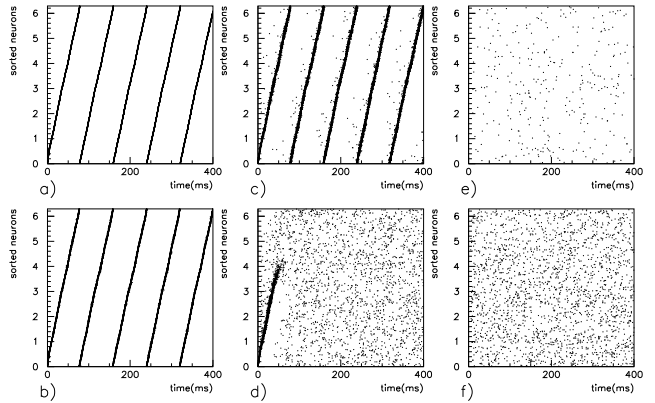}
\end{center}
\caption{
(a,b,c,d) Robustness wrt noise.
Raster plots show that, when the pattern retrieval is triggered,
network's spikes continue to have phase alignments resembling
the pattern even in presence of noise. Errors do not sum up until the system is
in the basin of attraction of the phase-coded pattern, as in a,b,c.
Different levels of noise  are used in a,b,c,d ($\sigma(J_{\text{noise}})=0,10,20,30$ respectively),
and pattern is triggered with $M=N/10$ as in previous cases.
Only in d the level of noise is too high and the system goes out of the basin of attraction.
(e,f) For comparison, the dynamics, when the retrieval is not triggered ($M=0$), is shown in
subplot e,f in presence of the same noise used in c,d. Figure e shows that the noise used in c 
usually affects strongly the dynamics of the network, however if the collective oscillation is retrieved the
 system is robust wrt noise. 
Thresholds in all figures are $\theta_{th}=80$, $N=3000$,
 and synaptic connections $J_{ij}$ are build learning $P=2$ phase-patterns at $\omega^\mu=3$Hz.
}
\label{noise}
\end{figure}

The network dynamics during the retrieval of a pattern in presence of noise is shown 
in Fig. \ref{noise} with different levels of noise
($\tau_{\text{noise}}=10$ms, $\bar J=0 $ and $\sigma(J_{\text{noise}})=0,10,20,30$ in a,b,c,d respectively).
Results show that when the noise is not able to move the dynamics out of the basin of attraction,
the errors do not sum up, and the phase relationship is preserved over time (see  Fig. \ref{noise}a,b,c).
If the input noise is very high, as in the example of  Fig. \ref{noise}d,  the dynamics moves out of the basin of attraction.\\
In order to see the effects of input noise level used in Fig \ref{noise}cd, 
we report  in Fig. \ref{noise}ef the network dynamics  when the pattern retrieval is not initiated ($M=0$). 
In particular, Fig. \ref{noise}e shows that the noise level used in Fig. \ref{noise}c is strong enough to generate spontaneous 
random activity in absence of the initial triggering, but is not sufficient to destroy the attractive dynamics during a successful
retrieval.
As in the Hopfield model, errors do not sum up and the dynamics spontaneously goes back  to the retrieved phase-coded pattern
for all  the  perturbations that leave the system inside the basin of attraction.

\begin{figure}[ht]
\begin{center}
a)
\includegraphics[width=3.8cm]{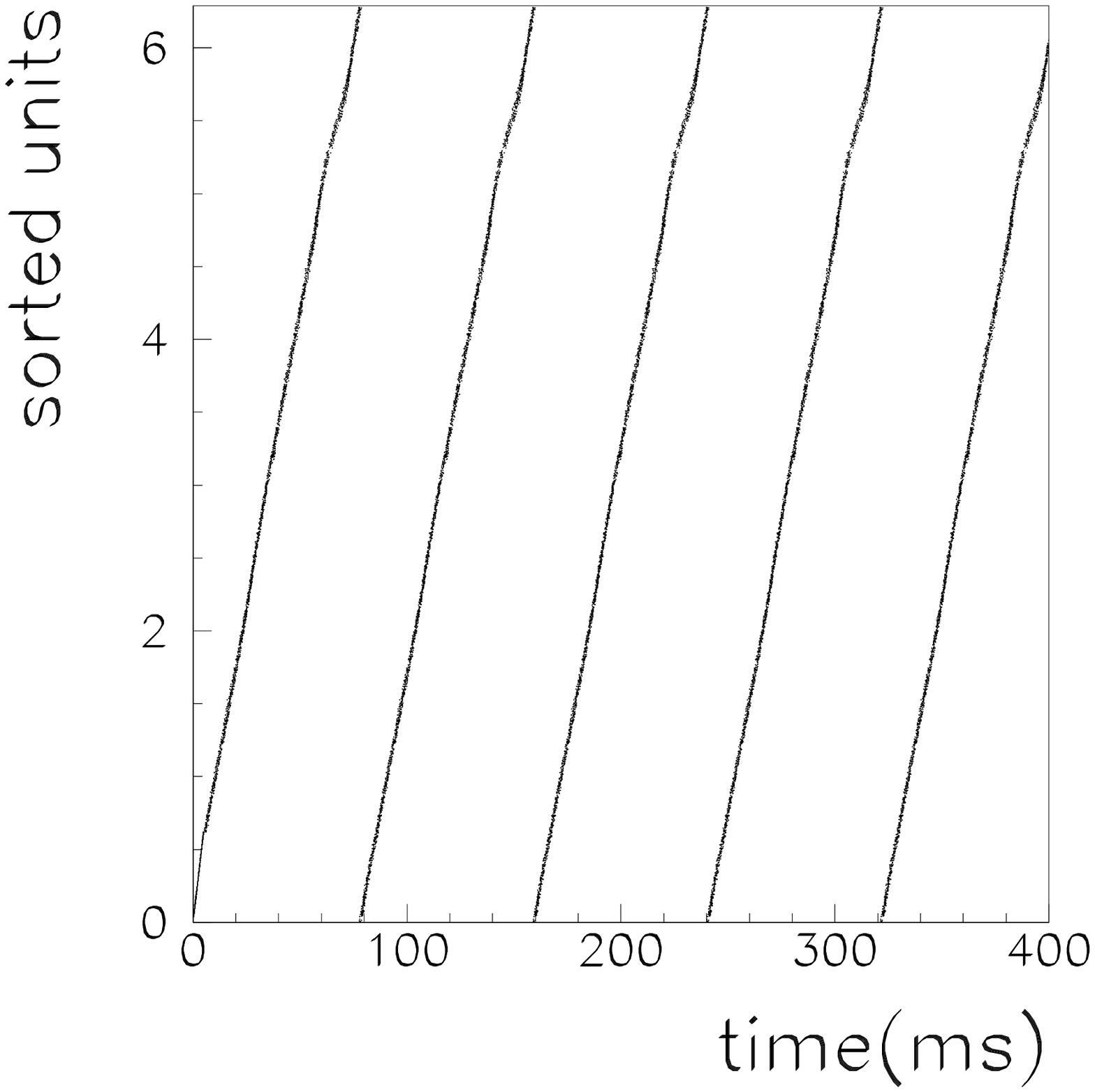}
b)
\includegraphics[width=3.8cm]{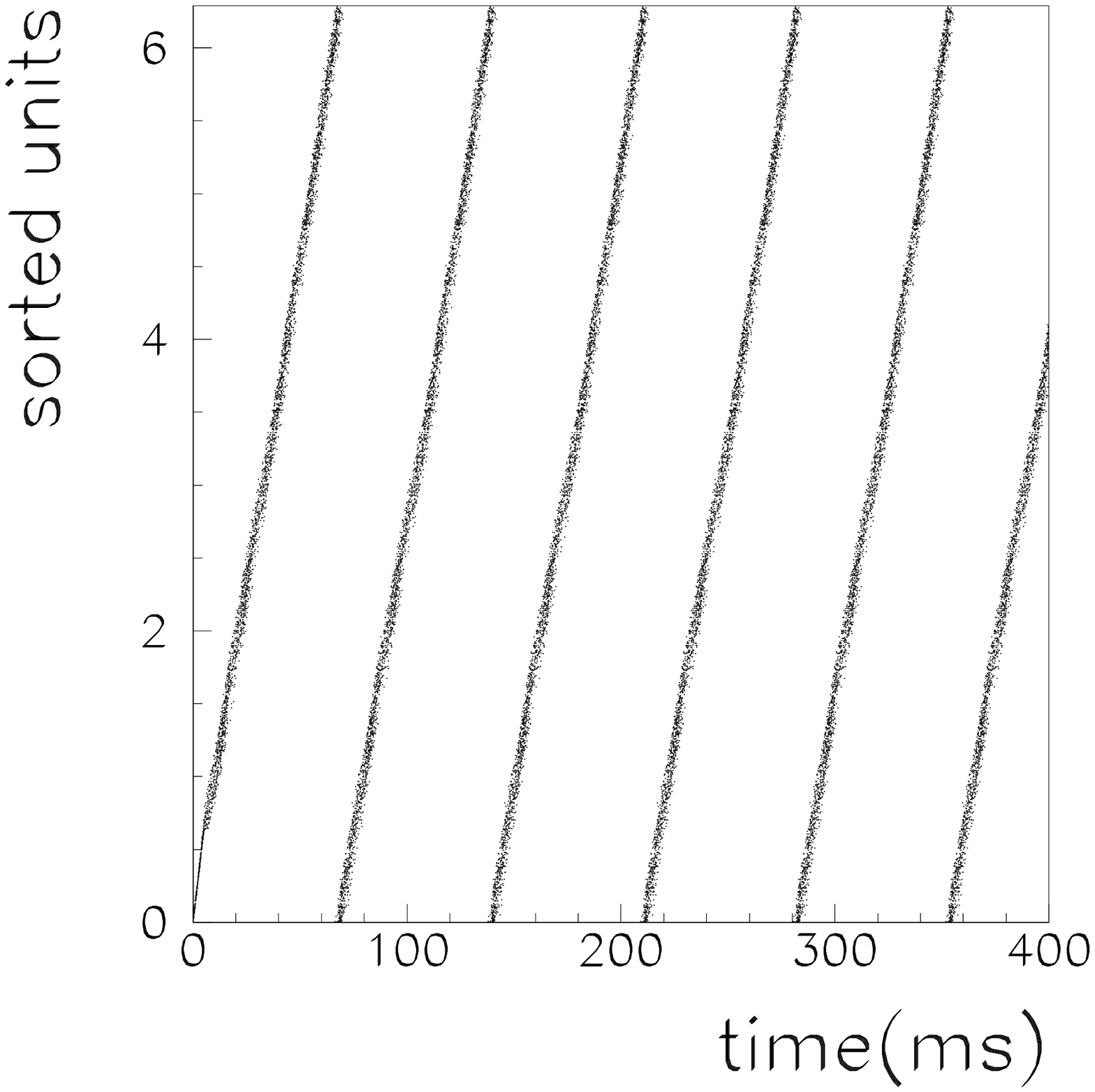}
\end{center}
\caption{
 Robustness wrt heterogeneity of spiking threshold values.
Raster plots show that, when the pattern retrieval is triggered,
units participate to the network collective oscillation, showing all the same frequency and  phase alignments resembling
the pattern, even in presence of threshold values heterogeneity among units.
Spiking thresholds of neuron $i$ are distributed according to $\theta_{th}^i=\theta_{th} (1+z \zeta)$ with average $\theta_{th}=80$ and
$z=0.2,0.5$ in (a,b) respectively. All other parameters are as in Fig.8a ( N=3000, M=N/10, and 
synaptic connections $J_{ij}$ are build learning $P=2$ phase-patterns at $\omega^\mu=3$Hz).
}
\label{noise-th}
\end{figure}

Lastly, the robustness of retrieval w.r.t.  heterogeneity of the spiking thresholds  is investigated.
This analysis can be carried out by using a different value $\theta_{th}^i$ of spiking threshold for each neuron $i$:
\begin{equation}
\theta_{th}^i= (1 + z \zeta_i) \theta_{th}
\end{equation}
where $\zeta_i$ is a random number extracted from a uniform distribution in $[-1,1]$,
and $z$ is the degree of heterogeneity.
Even with high degree of heterogeneity,
 the emergence of the retrieval collective dynamics  
forces all neurons to have exactly the same frequency of
oscillation and to keep a precise phase relationship, in a very robust manner. 
Figure  \ref{noise-th}a,b shows the dynamics  
with threshold heterogeneity  $z=0.2, 0.5$  respectively,
while the remaining parameters are set to the values of Fig. \ref{noise}a.\\
The above analysis shows that a unique  collective frequency  emerges, which is the 
frequency corresponding to the mean value of the $\theta_{th}^i$.
This is also evident by comparing Fig. \ref{noise-th}a,b with  Fig. \ref{noise}a. 
At $z=0.5$ it can be seen an additional small phase shift proportional to the value of $\theta$ of each neuron, but
collective activity is preserved.
Clearly if $z$ is too high and threshold values are distributed out of the region of successful retrieval
the network is unable to retrieve the pattern and failure  happens, as already discussed in Sec. \ref{sec-capacity}.

\section{Relationship of this model with theories of path-integration}
\label{sec-oscillators}

Recently,  path-integration system and the hippocampal-entorinal cortex circuit
have been deeply investigated \cite{Okeefe,Burgess,Burgess-2,Blair2008,Bruce,GeislerPNAS,Moser}, showing that place cells and grid cells form a map in which 
precise phase relationship among units play a central role to generate the spatial tuning.
A number of models of the spatial firing properties of place and grid cells were offered. 
Generally two main categories are distinguished: models which focus on continuous attractor mechanisms and models which use  interference
between oscillators at dynamically modulated frequencies; however, deeper computational
principles may exist that unify the different cases of neural integration \cite{pnas12}. 
In oscillatory interference models \cite{Burgess,Burgess-2,Blair2008,Welday2011,Zilli2010} (see also
\cite{moser11} for a review)
the total synaptic input to a neuron
(such as a grid cell or a place cell) is a weighted sum of the activities of n
oscillatory inputs,   whose oscillation frequency
is modulated by the rat velocity and head direction. 
Grid and place cells, according to these models,
derive their temporal and spatial  properties simply by
detecting synchrony among such velocity-modulated oscillatory inputs.
The oscillatory
interference theory is one possible hypothesized mechanism of path
integration and it has not been conclusively accepted or
rejected. It is  supported by recent studies suggesting that the predicted velocity-modulated oscillators 
exist as “theta cells” (interneurons found throughout the septo-hippocampal circuit) whose inter-burst frequency shows a cosine modulation by running direction and a linear increase with running velocity \cite{Welday2011}.

Such velocity-modulated oscillatory input was hypothesized to come from single oscillatory neurons \cite{Burgess}, or  networks such as 
subcortical "ring attractors" generating velocity-modulated theta oscillations\cite{Blair2008,Zilli2010,Welday2011}.

The properties of modulation and stability of frequency,
 and stability of multiple phase relationships,
 make our circuit a possible mechanism to build the  
velocity-modulated oscillators of the oscillatory interference theory. 
Indeed our circuit has a collective oscillation frequency,  which 
depends on the frequency stored in the  connectivity matrix, and that can 
be modulated 
 by changing the parameters such as $\theta_{th}$. 
Each neuron in the circuit has a phase determined by its position in that network,
i.e. determined by the phase $\phi_i^\mu$ of the stored  phase-pattern. 
If the parameter $\theta_{th}$ is modulated by animal speed, then the collective oscillation frequency of the circuit 
 is modulated by the animal's speed, while the neurons preserve stable phase relationship
among them.
\\
The other persistent-firing models \cite{Blair2008,Zilli2010,Welday2011} of the oscillators needed
by the oscillatory interference theory suffer from problems related to robustness, 
as those encountered by the single-cell oscillatory models \cite{Zilli}, 
due to the variability in the frequency of persisting spiking \cite{moser11}.
Indeed the oscillatory interference models impose strict constraints
upon the dynamical properties of the velocity-modulated oscillatory inputs, which have to preserve robust velocity-modulated frequency and
stable phase relationships among them on
relevant time scales in a manner robust to noise (many seconds, or dozens of theta cycle    
periods) \cite{Blair2008,Fiete,Zilli,moser11}.  
\\
Notably, in our model, since connections $J_{ij}$ among units in the circuit are fixed by the learning rule (\ref{conn}),
dynamics is a retrieval process and neurons preserve stable precise phase relationship among units and stable frequency
even in noisy conditions, at least when the dynamics is in the basin of attraction of that phase relationship.

Even in  the more recent spiking models \cite{Zilli2010,Welday2011}
of the oscillatory interference principle, in which many problems related to noise are solved, 
the heterogeneity of parameters of the cells which participate to the ring oscillator is not taken
into account.
\\
Here we show that the circuit level interactions among units make
the oscillation frequency and the phase-relationship of the system robust 
even with respect to heterogeneity of the spiking thresholds of the
units (see Sec. \ref{sec-noise}) .

This robustness, due to the proposed coupling which forces all the units of the circuit to have exactly the same period of oscillation
and to have precisely the same phase relationships of stored pattern, may be useful in all cases of sequence coding.

Moreover, our circuit can be easily programmed to cycle
 in different phase orders, by storing more than one phase-pattern as attractors.
The circuit  is a robust  “phase-shuffling ring oscillator”,  since the network has
the capability of ‘shuffling’ the order in which its neurons fire, by storing a variety of
different  phase-patterns within the connectivity. 
 If the oscillators predicted by the interference theory can generate more than one phase sequence, as
in the model presented here, then this could provide a potential mechanism
to explain the phenomenon of hippocampus  remapping\cite{Moser1,Cacucci}.  
One of the more interesting discovery of place cells behavior is indeed the
‘remapping’ of the place cell representation of space in response to
a changes in sensory or cognitive inputs, i.e. place cells change their
firing properties (place cells can appear disappear
or move to other unpredictable locations). This change may be abrupt and 
similar to the switch from one attractor to another\cite{Cacucci}.
If the place cell will fire
at a specific ‘place’ where its inputs become synchronized\cite{Bruce,Blair2008,Welday2011,moser11},
by recalling a different phase-coded pattern among the ones stored in 
 our circuit, it will change the phases of theta cells that are 
the inputs of the place cell, and it will change the specific 'place' where the inputs become

Finally we note that even thou our model is not a continuous attractor,  it shares many similarity with such a class of models.
Our model is a circuit with many distinct attractors, one for each phase-relationship stored in the
network, and the number of different attractor states is set by the maximal storage capacity studied here.
Furthermore each attractor is a phase-coded pattern, replayed with a collective resonant frequency that can be modulated
by changing for example the spiking threshold of the units.

Even thou during exploration the activity of place cells  may be explained  by the superposition of velocity-controlled oscillators inputs,
the recurrent connections inside the place cells network 
 may have anyway a relevant role.

During sleep, in absence of external input, the role of recurrent connections increases, probably due to an increase of excitability via neuromodulators or other mechanisms, and the spontaneous activity of the network show temporarily short replay of stored patterns, probably initiated simply by noise.

So the pattern activated repeatly during experience, is stored in the connectivity, and then activated during sleep when the network is near a critic point and noise is able to initiate short replay sequences.

Replay of phase-coded patterns of neural activity during sleep has been observed in hippocampus and neocortex\cite{Bruce2011}

Notably in our model the time scale of reactivation is different from the time scale of storing, depends on the
collective frequency which emerges from the connectivity, and may be accelerated or slowed down changing parameters such as spiking thresholds.
Therefore, our model might be also  relevant for replay in prefrontal cortex (PFC) or other cortical areas in which replay is accelerated
with respect to awake activity.

In the hippocampus, spikes representing adjacent place fields occur in rapid succession within a single theta cycle during behavior.
Therefore,  relative to this within–theta cycle rate,
reactivation during sleep in hippocampus is not accelerated. 
However,  reactivation in rat PFC is clearly compressed five to eight times  relative to the waking state\cite{Bruce,Bruce2011}. 
Indeed, while in hippocampus one may think that the coding sequence is the within-theta cycle, in prefrontal cortex
it is clearly seen that the cross-correlation among cells during sleep replay is time compressed compared with the cross-correlation during waking state. The playback speed declines over time as does the strength of the replay,  which is consistent for example
 with a simple increases of spiking threshold in our model.

\section{Discussion }
\label{sec-summary}
 
We studied the temporal dynamics, including the storage and replay properties, in a network of spiking
integrate and fire neurons, whose learning mechanism is based on the Spike-Timing
Dependent Plasticity.
The temporal patterns we consider are periodic spike-timing sequences,
whose features
are encoded in the relative phase shifts between neurons.

The importance of oscillations and precise temporal patterns has been pointed
out in many brain structure, such as cortex \cite{osc1}, cerebellum \cite{solinas,egidiodangelo}, or olfactory system \cite{osc2}.
The proposed associative memory approach, with selective replay of stored
sequence, can be a method for recognize an item, by activating the
same memorized pattern in response to a similar input. 
Another possibility is to have a way to transfer a memorized item to another area of the
brain, such as for memory consolidation during sleep.
During sleep, indeed, few spikes with the right phase relationship may initiate the
retrieval of one of the patterns stored in the network and
this reactivation may be useful for memory consolidation.
The stored pattern is an attractor of the network dynamics,
that is the dynamics spontaneously goes back to the  retrieved phase-coded pattern
for all  the  perturbations which leave the system within the basin of attraction.
Therefore  phase errors do not sum up, and the phase relationships  may be transferred 
and kept stable  over long time scales.\\
The time scale of the pattern during retrieval, i.e. the period of oscillation $1/\nu$,
  depends on (1) the  time constants of the single neuron $\tau_m$ and $\tau_s$, (2) the spiking threshold
$\theta_{th}$ of the neurons, and (3) the connectivity, through the STDP learning shape $A(\tau)$
and the time scale of the pattern during learning mode $1/\nu^\mu$.
Different areas of the brain may have different shape of STDP to subserve different oscillation frequencies
and different functional role. Here we fix the shape of $A(\tau)$ to the one observed in hippocampal 
cultures  (see Fig. 1 and \cite{biandpoo}) and   focus  on the dependence on the spiking threshold
$\theta_{th}$ of the neurons. The spiking threshold can modulate the
frequency of the collective oscillation, leaving unaffected the phase relationships among the units.
This opens a possible  way to govern the frequency of collective oscillation 
via neuromodulators, and to encode information (such as velocity of the animal)
 in the frequency of the oscillations, in addition to the information encoded in the phase relationships.
Notably, in a particular range of frequencies, the spiking threshold does not affect  
the frequency of oscillation but changes the number of spike per cycle during the
retrieval dynamics. 
This means that  information can be encoded via the number of spikes per cycle, 
independently from the information coded in the phase relationship among units,
in agreement with the observations of independent rate and phase coding 
in hippocampus  \cite{huxter_burgess}. 
Important to note, the phase relationships and the frequency of the collective oscillation are both robust with respect
to noise and to heterogeneity of the spiking threshold of the units.\\
A systematic study of the retrieval capacity of the network is proposed as a function of  two parameters of
the model: the frequency of the input pattern and the spiking threshold.
The storage capacity, evaluated as $P_{max}/N$,  is always lower than the storage capacity of the Hopfield model. 
However, the information content of a single pattern  in our dynamical model with N units
is higher than the information content of a pattern in the Hopfield model with N-units.
Indeed, an Hopfield pattern is a set of N binary 
values while our phase-coded pattern is a set of N real number $\phi^\mu_j \in [0, 2\pi]$.\\
The role of STDP  in the formation of  sequences
 has been recently investigated in \cite{newSERGIO,newFIETE}. These studies
have shown how it's possible to form long and complex sequences , but they
 did not concern themselves on how it's possible to learn and store
 not only the order of activation in a sequence, but  the precise
relative times between
 spikes in a closed sequence, i.e. a phase-coded pattern.
In our model not only the order of activation is preserved, but also the precise phase relationship among units.
The tendency to synchronization of units  is avoided in our model,
 without need to introduce delays or adaptation, due to the
 balance between excitation and inhibition that is in the connectivity
of large networks
 when  the phase coded pattern with random phases  is learned using
the rule in eqs. (5-8).
 In our rule all the connections, both positive and negative, scale
with the time of presentation
of patterns, keeping always a balance.
 Indeed, since (1) the stored phase are uniformly distributed in $[0,2\pi)$ 
and   (2) the learning window has the property  $\int A(\tau) dt = 0$,
then the connectivity matrix in eq.\ (7) has
the property that the summed excitation $(1/N)\sum_{i ,\\J_{ij}>0} J_ {ij}$
and the summed inhibition  $(1/N)\sum_{i ,\\J_{ij}<0} J_ {ij}$
are equal in the thermodynamic limit (indeed they are of order unity, while their difference
is of order $1/\sqrt{N}$). 

Under this conditions, we investigate how multiple phase-coded
patterns can be learned and selectively retrieved in the same network,
as a function of time scale of patterns and network parameters.

The task of storing and recalling phase-coded memories has been also
investigated in \cite{Mate05} in the framework of probabilistic
inference. While we study the effects of couplings given by Eqn.(5) in a network of IF neurons, the
paper \cite{Mate05} studies this problem from a normative theory
of autoassociative memory,  
in which real variable $x_i$ of neuron $i$ represents the
neuron $i$ spike timing with respect to a reference point of an ongoing field
potential, and the interaction $H(x_i, x_j)$ among units is mediated by the
derivative of the synaptic plasticity rule used to store memories.

The model proposed here is a mechanism which combines oscillatory and attractor dynamics, 
which may be useful in many models of path-integration, as pointed out in sec. (\ref{sec-oscillators}). Our learning model 
offers a IF circuit able to keep robust phase-relationship among cells participating
to a collective oscillation,  with a modulated collective frequency, robust with respect to noise and heterogeneities.
Notably the frequency of the collective oscillation in our circuit is not sensible to the single value of
the threshold of each unit, but to the average value of the threshold of all units, since all units participate to a
single collective oscillating pattern which is an attractor of the dynamics.

Recently there is renewed interest in reverberatory activity\cite{bi-reverbe} and 
in  cortical spontaneous activity\cite{spont,Luczak} 
whose spatiotemporal
structure  seems to reflect
the underlying connectivity, 
 which in turn may be the result of the past experience stored in the connectivity.\\ 
Similarity between spontaneous and evoked cortical activities has been shown to increase with age 
\cite{Mate11}, and with repetitive presentation of the stimulus \cite{reverberation}.
Interestingly, in our IF model, in order to induce spontaneous patterns of activity reminiscent of those stored during learning stage,
few spikes with the right phase relationship are sufficient.
It means that,  even  in absence of sensory stimulus, 
a noise with the right phase relationships  may induce a pattern of activity reminiscent of a stored pattern.
Therefore, by adapting the network connectivity to the phase-coded patterns observed during the learning mode,
the network dynamics builds a representation of the environment and  is able to replay the patterns of activity 
when stimulated by sense  or by chance.

This mechanism of learning phase-coded patterns of activity
 is then a way to  adapt the internal connectivity such that the network dynamics have attractors which
represent the patterns of activity seen during experience of environment.

\end{document}